\documentclass[twocolumn]{autart}

\usepackage{graphicx}
\usepackage{amsmath,amssymb,amsfonts}

\usepackage{booktabs}
\usepackage{url}
\usepackage{epstopdf}
\usepackage[authoryear]{natbib}
\usepackage{stmaryrd}
\usepackage{xcolor}

\newcommand{\isdef}{\stackrel{\Delta}{=}}
\newenvironment{prff}{\textbf{Proof}. }{}

\theoremstyle{plain}
\newtheorem{theorem}{Theorem}[section]

\newtheorem{corollary}[theorem]{Corollary}
\theoremstyle{definition}

\newtheorem{assumption}[theorem]{Assumption}
\theoremstyle{remark}

\newtheorem{examples}{\indent Example}[section]

\usepackage{xcolor}
\begin{document}
\begin{frontmatter}

\title{Finite-Time Markov-Parameter Identification of LTI Systems Using
Non-Causal FIR Models: A Unified Framework for Stable and Unstable Systems}

\author[VT]{Ahmad Al-Tawaha}\ead{atawaha@vt.edu}
\author[VT]{Ming Jin}
\author[JUST]{Khaled F. Aljanaideh}

\address[VT]{Bradley Department of Electrical and Computer Engineering, Virginia Tech, Blacksburg, VA, USA}
\address[JUST]{Department of Aeronautical Engineering, Faculty of Engineering, Jordan University of Science and Technology, Irbid, Jordan}

\begin{keyword}
Closed-loop identification; Unstable LTI systems; Non-causal FIR; Instrumental variables.
\end{keyword}                            

\begin{abstract}
We present a finite-time framework for identifying stable and unstable linear time-invariant (LTI) systems from a single closed-loop input-output trajectory. The method does not require knowledge of the stabilizing controller, an intermediate observer, or prior separation of the plant into stable and unstable components. The approach uses a non-causal finite impulse response (FIR) model obtained from a Laurent expansion of the transfer function. In this representation, stable dynamics are captured by causal Markov parameters, while unstable dynamics are captured by non-causal coefficients associated with reverse-time stable evolution. This avoids the growth of causal unstable Markov parameters. A key advantage is that the coefficients multiplying both the input and the process noise remain controlled by stable and reverse-time stable decay rates, rather than by growing forward-time unstable dynamics. To handle closed-loop data, we use the injected excitation as an instrumental variable, which removes the bias caused by correlation between the feedback input and the process noise. Under explicit instrument-strength and closed-loop concentration conditions, we derive a non-asymptotic error bound for the estimated Laurent/FIR Markov parameters with the usual $\mathcal{O}(N^{-1/2})$ statistical rate, up to logarithmic factors and truncation terms. The bound captures the effects of process noise, measurement noise, FIR horizons, closed-loop state moments, and controller-dependent instrument conditioning. Numerical experiments support the finite-time analysis by showing the predicted Markov-parameter convergence rate and illustrating how controller-dependent instrument conditioning affects the sample complexity of closed-loop identification.
\end{abstract}

\end{frontmatter}

\section{Introduction}
Recent years have witnessed a surge of interest in the machine learning community in data-driven approaches to control and system identification, particularly for learning unknown dynamical systems from data \citep{regularizationrobust,sarkar2021finite,lale2021adaptive,lee2020non}. Linear dynamical systems (LDS) underpin numerous real-world processes ranging from robotics and industrial automation to forecasting time-series data such as financial and climate patterns \citep{stokey1989recursive,goodwin2014adaptive,brunton2016discovering,ng2006autonomous}. Unlike classical approaches that provide only asymptotic convergence guarantees \citep{ljung1998system,campi1998adaptive}, many practical scenarios---such as large-scale power networks or robotic systems---require finite-time performance guarantees because collecting extensive data or multiple independent trajectories can be prohibitively costly \citep{fattahi2019learning,tu2019sample}. Consequently, recent attention has shifted towards non-asymptotic analysis, emphasizing the relationship between estimation accuracy and finite-sample complexity \citep{simchowitz2018learning,sarkar2019near,dean2020sample}. While finite-time guarantees exist for stable systems with partial observability \citep{lee2022improved,oymak2021revisiting,simchowitz2019learning,tsiamis2019finite,sarkar2021finite,djehiche2022efficient,fattahi2021learning} and for unstable systems with full observability \citep{faradonbeh2018finite,sarkar2019near}, identifying \textit{partially observed and potentially unstable LTI systems from a single, finite-length trajectory} remains a significant open challenge of profound importance for developing robust, real-world control applications.

Modern control systems predominantly operate in closed-loop configurations, making it impractical or unsafe to open the loop for system identification purposes \citep{hjalmarsson1996model,landau2001identification}. This constraint is particularly critical for open-loop unstable systems, where safety considerations strictly prohibit disabling feedback control. Existing closed-loop identification methods with finite-time guarantees impose restrictive assumptions. Approaches like \citep{lale2020logarithmic,lale2021adaptive,jones2022closed,lale2021finite,regularizationrobust} employ observers or predictor-based formulations but assume \textit{open-loop stability}. Even methods that address unstable systems introduce alternative constraints: \citep{szentpeteri2023non} requires full observability of states, \citep{lee2020non} demands a \textit{known} linear controller while focusing on output estimation rather than model recovery, and \citep{jones2022closed} relies on a known linear feedback structure. Prior results such as \citep{oymak2019non} can, in principle, be extended to unstable plants under a stabilizing controller, but this extension requires \textit{knowledge of the controller policy} to decorrelate the closed-loop data---a requirement that the proposed approach eliminates.

Additionally, most prior works rely on autoregressive
models with exogenous inputs (ARX or VARX)
\citep{lale2020logarithmic,lale2021adaptive,jones2022closed,
lale2021finite}. These approaches impose notable constraints:
they require prior knowledge of system order or stability
characteristics, and exhibit unfavorable sample complexity that
scales polynomially with system dimension and past horizon length.

Prior non-causal FIR approaches established the usefulness of
Laurent representations for closed-loop identification of unstable
systems \citep{aljanaideh2017closed}. In contrast, our focus is a
finite-sample IV analysis of the estimated Laurent/FIR Markov
parameters from a single closed-loop trajectory without knowledge of
the stabilizing controller.

A fundamental difficulty with unstable systems in causal finite-sample identification analyses is that causal Markov parameters and the coefficients multiplying process noise can involve growing forward-time powers of the unstable dynamics \citep{oymak2019non,simchowitz2018learning,sarkar2019near, sarkar2021finite,tsiamis2019finite}. Motivated by earlier non-causal FIR and Laurent-series approaches, we use a two-sided Laurent/FIR representation in which the unstable component is described through reverse-time stable dynamics. The Laurent coefficients multiplying both the input and the process noise are therefore governed by stable and reverse-time stable decay rates, rather than by growing forward-time unstable dynamics. The resulting bounds remain controlled up to transient-amplification and realization-conditioning constants.

\textbf{Contributions.} We summarize the main contributions below.
\begin{itemize}

\item \textbf{Closed-loop non-causal FIR identification without controller knowledge.} We develop a non-causal Laurent/FIR framework for estimating the Markov parameters of stable and unstable LTI systems from a single closed-loop trajectory. The learner uses the measured input and output together with the known injected excitation, but does not require full-state observations, knowledge of the stabilizing controller, an intermediate observer, or a prior stable and unstable decomposition of the plant. The injected excitation is used as an instrumental variable to remove the bias caused by closed-loop correlation between the feedback input and the process noise.

\item \textbf{Finite-sample guarantees with controlled unstable and process-noise terms.} We prove finite-sample error bounds for the estimated Laurent/FIR Markov parameters, with an $\mathcal{O}(N^{-1/2})$ statistical rate up to logarithmic factors and truncation terms. The non-causal representation captures unstable dynamics through reverse-time stable coefficients, so the terms multiplying both the input and the process noise are controlled by stable and reverse-time stable decay rates rather than by growing forward-time unstable dynamics. The resulting bound separates the effects of process noise, measurement noise, truncation tails, and instrument conditioning.

\item \textbf{Controller-dependent sample complexity and recursive implementation.} The analysis makes explicit how the unknown stabilizing controller affects sample complexity through closed-loop state moments, empirical concentration of the feedback and instrument cross-covariance, and the strength of the injected excitation as an instrument. This shows that a controller may stabilize the plant while still producing weak instrument conditioning and requiring more samples. The proposed formulation also supports recursive updates of the non-causal FIR coefficients with a fixed $d$-sample delay; for closed-loop data, the recursive IV update uses the empirical cross-covariance between the input regressor and the instrument regressor.
\end{itemize}

\vspace{-10pt}
\section{Background and Preliminaries}
\vspace{-5pt}
\label{back}

Consider a discrete-time linear time-invariant (LTI) system
\begin{equation}
\begin{aligned}
x(k+1) &= Ax(k) + Bu(k) + B_{w}w(k), \\
y(k) &= Cx(k) + Du(k) + v(k),
\end{aligned}
\label{ssEqn}
\end{equation}
where $x(k)\in\mathbb{R}^{n}$, $u(k)\in\mathbb{R}^{p}$,
$y(k)\in\mathbb{R}^{m}$, $w(k)\in\mathbb{R}^{l}$,
$v(k)\in\mathbb{R}^{m}$,
$A\in\mathbb{R}^{n\times n}$, $B\in\mathbb{R}^{n\times p}$,
$B_{w}\in\mathbb{R}^{n\times l}$,
$C\in\mathbb{R}^{m\times n}$, and
$D\in\mathbb{R}^{m\times p}$.
The learner observes a single trajectory $\{y(k),u(k),c(k)\}_{k=0}^{\ell}$, where $\ell+1$ is the number of samples and $c(k)$ is the known injected excitation used as the instrumental variable. The system matrices are unknown to the learner.

Because open-loop experiments with unstable plants may be unsafe,
we consider data collected under feedback. The control input at time
step $k$ takes the form
\begin{equation}
f(k)
\isdef
\mathcal{K}\!\left(y(0),\ldots,y(k-1)\right),
\quad
u(k)=f(k)+c(k),
\label{eq:feedback_component}
\end{equation}
where $\mathcal{K}$ is a \textit{strictly causal}, possibly nonlinear
controller that is \textit{unknown} to the learner, and
$c(k)\in\mathbb{R}^{p}$ is a known excitation signal injected on top of the feedback action. The notation $\mathcal{K}(y(0),\ldots,y(k-1))$ means that, at time $k$, the controller maps the finite output history $(y(0),\ldots,y(k-1))$ to a vector in $\mathbb{R}^{p}$. For $k=0$, the controller acts on the empty output history. Strict causality means that $\mathcal{K}$ depends on outputs only up to time step $k-1$ and therefore cannot depend on the current output $y(k)$ or on the current injected excitation $c(k)$ through $y(k)$.


\begin{assumption}
\rm
\label{ass:closedloop}
The feedback interconnection defined by~\eqref{ssEqn} and
\eqref{eq:feedback_component} is well posed and mean-square stable. In
particular,
\begin{equation}
\sup_{k\ge0}\mathbb{E}\|x(k)\|^{2}<\infty .
\label{eq:mean_square_stability}
\end{equation}
\end{assumption}


\begin{assumption}
\rm
\label{ass:excitation}
The sequences $\{c(k)\}$, $\{w(k)\}$, and $\{v(k)\}$ are mutually
independent i.i.d.\ Gaussian such that
\[
c(k)\sim\mathcal{N}(0,\sigma_{c}^{2}I_{p}),
w(k)\sim\mathcal{N}(0,\sigma_{w}^{2}I_{l}),
v(k)\sim\mathcal{N}(0,\sigma_{v}^{2}I_{m}),
\]
and the initial state satisfies $x(0)=0$.
\end{assumption}

\begin{assumption}
\rm
\label{assumptionMinimal}
$(A,B,C)$ is a minimal realization of
$G(z)=C(zI-A)^{-1}B+D$.
\end{assumption}

\begin{assumption}
\rm
\label{ass:no_poles}
The transfer matrix $G(z)=C(zI-A)^{-1}B+D$ has no poles on the unit
circle.
\end{assumption}

By Assumption~\ref{ass:closedloop}, the closed-loop state has uniformly bounded second moments. We denote this bound by
\begin{equation}
\Gamma_{\mathrm{cl}}
\isdef
\sup_{k\ge0}
\left\|
\mathbb{E}\!\left[x(k)x(k)^{\top}\right]
\right\|
<\infty .
\label{eq:Gamma_cl_bound}
\end{equation}

\vspace{-5pt}
\subsection{Non-causal FIR representation via Laurent expansion}

Under Assumptions~\ref{assumptionMinimal} and~\ref{ass:no_poles},
$G$ can be uniquely decomposed as
\begin{equation}
G(z)=G_{\mathrm{s}}(z)+G_{\mathrm{u}}(z)+D,
\label{G_decomp}
\end{equation}
where $D\isdef G(\infty)$, $G_{\mathrm{s}}$ is the strictly proper
stable part, whose poles lie inside the open unit disk, and
$G_{\mathrm{u}}$ is the strictly proper unstable part, whose poles lie
outside the closed unit disk.

Equivalently, after a similarity transformation, the realization can
be written as
\begin{equation}
\begin{aligned}
A
&=
\begin{bmatrix}
A_{\mathrm{s}} & 0\\
0 & A_{\mathrm{u}}
\end{bmatrix},
&
B
&=
\begin{bmatrix}
B_{\mathrm{s}}^{\top} & B_{\mathrm{u}}^{\top}
\end{bmatrix}^{\top},
\\
B_w
&=
\begin{bmatrix}
B_{\mathrm{s},w}^{\top} & B_{\mathrm{u},w}^{\top}
\end{bmatrix}^{\top},
&
C
&=
\begin{bmatrix}
C_{\mathrm{s}} & C_{\mathrm{u}}
\end{bmatrix}.
\end{aligned}
\label{eq:stable_unstable_realization}
\end{equation}
where $\rho(A_{\mathrm{s}})<1$, $\rho(A_{\mathrm{u}}^{-1})<1$, and $n_{\mathrm{s}}+n_{\mathrm{u}}=n$.

The key insight is that $G_{\mathrm{s}}(z)$ admits a causal power
series in $z^{-1}$, while $G_{\mathrm{u}}(z)$ admits a non-causal power
series in $z$. Together they form a Laurent series analytic in an open
annulus $\rho_{\mathrm{s}}<|z|<\rho_{\mathrm{u}}$ containing the unit
circle:
\begin{equation}
G(z)=\sum_{i=-\infty}^{\infty}H_{i}z^{-i},
\qquad
H_i=
\begin{cases}
C_{\mathrm{s}}A_{\mathrm{s}}^{i-1}B_{\mathrm{s}},
& i\ge 1,\\[1mm]
D-C_{\mathrm{u}}A_{\mathrm{u}}^{-1}B_{\mathrm{u}},
& i=0,\\[1mm]
-C_{\mathrm{u}}A_{\mathrm{u}}^{i-1}B_{\mathrm{u}},
& i\le -1.
\end{cases}
\label{LEGyu}
\end{equation}
For $i\ge1$, the positive-lag Markov parameters are
\[
H_i
=
C_{\mathrm{s}}A_{\mathrm{s}}^{i-1}B_{\mathrm{s}},
\]
so their norms decay geometrically with $i$ because
$\rho(A_{\mathrm{s}})<1$. For the negative-lag coefficients, let
$i=-j$ with $j\ge1$. Then
\[
H_{-j}
=
-C_{\mathrm{u}}A_{\mathrm{u}}^{-j-1}B_{\mathrm{u}}
=
-C_{\mathrm{u}}
(A_{\mathrm{u}}^{-1})^{j+1}
B_{\mathrm{u}}.
\]
Thus the negative-lag coefficients decay geometrically with $j$ because $\rho(A_{\mathrm{u}}^{-1})<1$. Hence, although $A_{\mathrm{u}}$ is unstable in forward time, its contribution to the Laurent series is governed by the stable reverse-time dynamics $A_{\mathrm{u}}^{-1}$. This is why the non-causal FIR representation can approximate unstable dynamics using bounded, decaying coefficients.

A non-causal FIR model is obtained by truncating the Laurent series
\eqref{LEGyu}:
\begin{equation}
G_{\mathrm{F},r,d}(z)
\isdef
\sum_{i=-d}^{r}H_i z^{-i}.
\label{GnonFIR}
\end{equation}

The Laurent expansion of the transfer function
$G_{y,w}(z)=C(zI-A)^{-1}B_w$ from the process noise $w$ to the output
$y$ is
\begin{equation}
G_{y,w}(z)=\sum_{i=-\infty}^{\infty}F_{i}\,z^{-i},\quad
F_i= \begin{cases}C_{\rm s} A_{\rm s}^{i-1}B_{{\rm s},w}, & i \geq 1, \\ -{C}_{\rm{u}}{A}_{\rm{u}}^{-1}B_{{\rm u},w}, & i=0, \\ -{C}_{\rm{u}}{A}_{\rm{u}}^{i-1}B_{{\rm u},w}, & i \leq-1.\end{cases}
\label{LEGyw}
\end{equation}
As above, the negative-lag coefficients contain powers of $A_{\mathrm{u}}^{-1}$ rather than growing powers of $A_{\mathrm{u}}$. Thus the process-noise coefficients are controlled by the reverse-time stable dynamics, up to realization-conditioning constants.

Using \eqref{ssEqn}, \eqref{LEGyu}, and \eqref{LEGyw}, the output can be written as
\begin{equation}
y(k)
=
\theta_{r,d}\phi_{r,d}(k)
+
\gamma_{r,d}\phi_{w,r,d}(k)
+
e_{r,d}(k)
+
v(k),
\label{yExpanded}
\end{equation}
where
\begin{align}
\theta_{r,d}
&\isdef
[H_{-d}\;\cdots\;H_{r}]
\in\mathbb{R}^{m\times p\mu},
\\
\mu
&\isdef
r+d+1,
\\
\phi_{r,d}(k)
&\isdef
[u(k+d)^{\top}\;\cdots\;u(k-r)^{\top}]^{\top}
\in\mathbb{R}^{p\mu},
\\
\gamma_{r,d}
&\isdef
[F_{-d}\;\cdots\;F_{r}]
\in\mathbb{R}^{m\times l\mu},
\\
\phi_{w,r,d}(k)
&\isdef
[w(k+d)^{\top}\;\cdots\;w(k-r)^{\top}]^{\top}
\in\mathbb{R}^{l\mu},
\end{align}
and
\begin{equation}
e_{r,d}(k)
\isdef
\underbrace{
C_{\mathrm{s}}A_{\mathrm{s}}^{r}x_{\mathrm{s}}(k-r)
}_{\text{causal tail}}
+
\underbrace{
C_{\mathrm{u}}A_{\mathrm{u}}^{-d-1}x_{\mathrm{u}}(k+d+1)
}_{\text{non-causal tail}} .
\label{truncationErrors2}
\end{equation}
The causal tail decays as $\rho(A_{\mathrm{s}})^r$ with the lookback horizon $r$, and the non-causal tail decays as $\rho(A_{\mathrm{u}}^{-1})^d$ with the preview $d$. The regressor $\phi_{w,r,d}(k)$ is used only for the analysis of the process-noise contribution; the estimator does not require observing $w(k)$.

The representation~\eqref{yExpanded} reduces identification to estimating the finite block of Laurent coefficients $\theta_{r,d}$. If the input were generated independently of the process noise, ordinary least squares could be applied directly to \eqref{yExpanded}. In closed loop, however, the input is generated from past outputs, and those outputs are affected by the process noise. Thus the regressor $\phi_{r,d}(k)$ can be correlated with the error terms in~\eqref{yExpanded}. The next subsection introduces an instrumental-variable estimator that removes this closed-loop bias by using the injected excitation $c$ as an instrument.

\vspace{-5pt}
%
\subsection{Instrumental-variable estimation for closed-loop data}
\label{sec:IV}

The goal of this subsection is to construct an estimator for $\theta_{r,d}$ that remains valid when the data are collected under feedback. The key observation is that the known excitation $c$ is independent of the process and measurement noises, but it is still correlated with the measured input $u$ through the identity $u=f+c$. Therefore, $c$ can be used as an instrument for the input regressor. Recall from \eqref{eq:feedback_component} that
\[
u(k)=f(k)+c(k),
\qquad
f(k)=\mathcal K(y(0),\ldots,y(k-1)).
\]
The feedback component $f(k)$ depends on past measured outputs and can
therefore be correlated with process noise through the closed-loop
dynamics. Consequently, ordinary least squares using the input regressor
$\phi_{r,d}(k)$ can be biased. To remove this closed-loop bias, we use
the injected excitation $c$ as an instrumental variable.

Define the instrument regressor
\begin{equation}
\phi_{c,r,d}(k)
\isdef
[c(k+d)^{\top}\;\cdots\;c(k-r)^{\top}]^{\top}
\in\mathbb{R}^{p\mu},
\label{eq:instrument_regressor}
\end{equation}
The feedback signal $f(k)$ is not an additional input observed
separately from $u(k)$. It is a conceptual decomposition of the
measured input into the unknown feedback action and the known injected
excitation. This decomposition is used to analyze the population
cross-covariance, while the estimator itself uses the observed
signals $u$, $y$, and $c$.
\begin{equation}
\phi_{f,r,d}(k)
\isdef
[f(k+d)^{\top}\;\cdots\;f(k-r)^{\top}]^{\top}
\in\mathbb{R}^{p\mu}.
\label{eq:feedback_regressor}
\end{equation}

Since $u(k)=f(k)+c(k)$, the FIR input regressor decomposes as
\begin{equation}
\phi_{r,d}(k)
=
\phi_{f,r,d}(k)
+
\phi_{c,r,d}(k).
\label{eq:regressor_decomposition}
\end{equation}

For $\ell=N+r+d-1$, define the data matrices
\begin{align}
\Psi_{y,\ell}
&\isdef
[y(r)\;\cdots\;y(\ell-d)]
\in\mathbb{R}^{m\times N},
\label{eq:Psi_y_def}
\\
\Phi_{r,d,\ell}
&\isdef
[\phi_{r,d}(r)\;\cdots\;\phi_{r,d}(\ell-d)]
\in\mathbb{R}^{p\mu\times N},
\label{eq:Phi_u_def}
\\
\Phi_{c,r,d,\ell}
&\isdef
[\phi_{c,r,d}(r)\;\cdots\;\phi_{c,r,d}(\ell-d)]
\in\mathbb{R}^{p\mu\times N},
\label{eq:Phi_c_def}
\\
\Phi_{f,r,d,\ell}
&\isdef
[\phi_{f,r,d}(r)\;\cdots\;\phi_{f,r,d}(\ell-d)]
\in\mathbb{R}^{p\mu\times N}.
\label{eq:Phi_f_def}
\end{align}
By \eqref{eq:regressor_decomposition},
\[
\Phi_{r,d,\ell}
=
\Phi_{f,r,d,\ell}
+
\Phi_{c,r,d,\ell}.
\]

The batch instrumental-variable estimator is
\begin{equation}
\hat{\theta}_{r,d,\ell}^{\mathrm{IV}}
=
\Psi_{y,\ell}\Phi_{c,r,d,\ell}^{\top}
\left(
\Phi_{r,d,\ell}\Phi_{c,r,d,\ell}^{\top}
\right)^{-1}.
\label{eq:IV_est}
\end{equation}

The IV estimator in~\eqref{eq:IV_est} normalizes by the empirical input and instrument cross-covariance $\Phi_{r,d,\ell}\Phi_{c,r,d,\ell}^{\top}$. Hence, the population counterpart of this matrix determines whether the instrument is informative enough to identify $\theta_{r,d}$. This motivates the following definition. Define the finite-horizon input and instrument cross-covariance
\begin{equation}
R_{uc}
\isdef
\mathbb E
\left[
\phi_{r,d}(k)\phi_{c,r,d}(k)^{\top}
\right]
\in\mathbb R^{p\mu\times p\mu}.
\label{eq:R_uc}
\end{equation}
Using \eqref{eq:regressor_decomposition}, we have
\begin{equation}
R_{uc}
=
\sigma_c^2 I_{p\mu}
+
S_{fc},
\quad
S_{fc}
\isdef
\mathbb E
\left[
\phi_{f,r,d}(k)\phi_{c,r,d}(k)^{\top}
\right].
\label{eq:Ruc_decomp}
\end{equation}
The first term comes from the direct excitation $c$ in $u=f+c$. The
second term captures how past injected excitations propagate through
the feedback loop and reappear in the feedback signal.

We now inspect the block structure of $R_{uc}$. The following indexing simply maps a block position in the non-causal regressor to the corresponding time instant.
\[
t_i(k)\isdef k+d+1-i,
\qquad
i=1,\ldots,\mu.
\]
Thus $t_1(k)=k+d$ corresponds to the first, most future, block of
the regressor, while $t_{\mu}(k)=k-r$ corresponds to the last, most
past, block. The $(i,j)$ block of $R_{uc}$ is
\[
[R_{uc}]_{ij}
=
\mathbb E
\left[
u(t_i(k))c(t_j(k))^{\top}
\right].
\]
The direct excitation term in $u=f+c$ contributes
$\sigma_c^2I_p$ when $i=j$ and zero otherwise. The remaining term is
\[
\mathbb E
\left[
f(t_i(k))c(t_j(k))^{\top}
\right].
\]
Because the controller is strictly causal, $f(t_i(k))$ depends only on
outputs before time $t_i(k)$. Those outputs may depend on past values
of $c$, but they cannot depend on $c(t_i(k))$ or on future values of
$c$. If $j\le i$, then $t_j(k)\ge t_i(k)$, so $c(t_j(k))$ is current or
future relative to $f(t_i(k))$. Therefore,
\[
\mathbb E
\left[
f(t_i(k))c(t_j(k))^{\top}
\right]
=
0
\qquad
\text{whenever } j\le i.
\]
When $j>i$, $t_j(k)<t_i(k)$, so $c(t_j(k))$ is a past excitation and
may influence $f(t_i(k))$ through the closed-loop dynamics. Hence the
feedback contribution can be nonzero only above the block diagonal.


Consequently, $S_{fc}$ is
block strictly upper triangular and
\begin{equation}
R_{uc}
=
\sigma_c^2
\left(
I_{p\mu}
+
\mathcal U
\right),
\label{eq:Ruc_triangular}
\end{equation}
where
\begin{equation}
[\mathcal U]_{ij}
=
\begin{cases}
\sigma_c^{-2}
\mathbb E
\left[
f(t_i(k))c(t_j(k))^{\top}
\right],
& j>i,\\[1mm]
0,
& j\le i.
\end{cases}
\label{eq:U_blocks}
\end{equation}
This triangular structure explains why strict causality is useful:
the diagonal blocks of $R_{uc}$ are fixed by the injected excitation,
while feedback affects only the upper-triangular off-diagonal blocks.
The matrix is therefore invertible at every finite horizon, but its
smallest singular value can still be small. This is why we impose a
quantitative conditioning assumption below.

\begin{assumption}
\rm
\label{ass:excitability}
The finite-horizon input and instrument cross-covariance is
quantitatively well conditioned. Specifically,
\begin{equation}
s_{\mathrm{IV}}
\isdef
\sigma_{\min}(R_{uc})>0.
\label{eq:siv_def}
\end{equation}
We also define the normalized instrument-strength parameter
\begin{equation}
\lambda_{\mathrm{IV}}
\isdef
\frac{s_{\mathrm{IV}}^2}{\sigma_c^2}
=
\frac{\sigma_{\min}(R_{uc})^2}{\sigma_c^2}.
\label{eq:lambda_iv_def}
\end{equation}
\end{assumption}

The normalized parameter $\lambda_{\mathrm{IV}}$ is used to express the
statistical error in a scale-invariant way. In the IV error identity,
the inverse cross-covariance contributes a factor proportional to
$1/s_{\mathrm{IV}}$, while products involving the instrument have scale
$\sigma_c$. Therefore the relevant ratio is
\[
\frac{\sigma_c}{s_{\mathrm{IV}}}
=
\frac{1}{\sqrt{\lambda_{\mathrm{IV}}}}.
\]
Small values of $\lambda_{\mathrm{IV}}$ correspond to weak instruments
and lead to larger estimation error.

\smallskip
\noindent
\textbf{Controller interpretations and instrument conditioning.}
The construction above only uses strict causality, the independence of
the injected excitation from the noise sequences, and the conditioning
of the finite-horizon cross-covariance $R_{uc}$. The following
observations clarify how the same definitions specialize to nonlinear,
linear, and open-loop settings.

\smallskip
\noindent
\textit{Nonlinear controllers.}
For a general strictly causal nonlinear controller, no differentiability
or impulse-response representation of $\mathcal K$ is assumed. The
blocks of $\mathcal U$ are defined directly by the cross-covariances in
\eqref{eq:U_blocks}. Strict causality gives the triangular population
structure because $f(t)$ cannot depend on $c(t)$ or on future values of
$c$. The additional difficulty for nonlinear controllers is
finite-sample concentration of the empirical feedback and instrument
cross-covariance. This is handled explicitly in
Assumption~\ref{ass:cl_concentration}.

\smallskip
\noindent
\textit{Linear closed-loop interpretation.}
The definition of $f(k)$ in~\eqref{eq:feedback_component} is the
general definition used throughout the paper. The triangular argument
above does not require a linear controller. It only uses strict
causality and the independence of the injected excitation from the noise
sequences.

When the closed-loop map from the exogenous signals to the feedback
action is linear and stable, the same feedback signal can be interpreted
through impulse responses. In particular, its component driven by the
injected excitation can be written as
\[
f_c(k)
=
\sum_{s=1}^{\infty}
\mathcal T_s^c c(k-s),
\]
where $\mathcal T_s^c$ is the closed-loop impulse response from the
injected excitation $c$ to the feedback action $f$. The remaining part
of $f(k)$ is driven by the process and measurement noises. These
noise-driven components do not contribute to
$\mathbb E[f(t)c(\tau)^{\top}]$ because $w$, $v$, and $c$ are mutually
independent. Therefore, in the linear case,
\[
\mathbb E[f(t)c(\tau)^{\top}]
=
\mathbb E[f_c(t)c(\tau)^{\top}].
\]
This linear impulse-response interpretation is useful for understanding
the off-diagonal blocks of $\mathcal U$, but it is not an additional
assumption in the IV estimator or in the finite-sample analysis below.

\smallskip
\noindent
\textit{A conservative conditioning bound.}
In the linear interpretation above, define
\[
\mathcal T_{\infty}
\isdef
\sum_{s=1}^{\infty}
\|\mathcal T_s^c\|.
\]
This quantity measures the total closed-loop gain from the injected
excitation $c$ to the feedback signal $f$. Since the strictly upper
triangular blocks of $\mathcal U$ are generated by these lagged
responses, one obtains the conservative bound
\[
\|\mathcal U\|
\le
\mathcal T_{\infty}.
\]
Thus, if $\mathcal T_{\infty}<1$, then
\[
s_{\mathrm{IV}}
=
\sigma_{\min}(R_{uc})
\ge
\sigma_c^2(1-\mathcal T_{\infty}),
\]
and consequently
\[
\lambda_{\mathrm{IV}}
\ge
\sigma_c^2(1-\mathcal T_{\infty})^2.
\]
This condition is only sufficient. It is not required by the IV
estimator or by the finite-sample analysis below. Large values of
$\mathcal T_{\infty}$ should be interpreted as an indication that the
instrument may be weakly conditioned, which increases the sample size
needed for accurate estimation.

\smallskip
\noindent
\textit{Open-loop stable case.}
If the system is open-loop stable and no feedback is used, then
$f(k)=0$ and $u(k)=c(k)$. Therefore,
\[
R_{uc}
=
\sigma_c^2I_{p\mu},
\qquad
s_{\mathrm{IV}}
=
\sigma_c^2,
\qquad
\lambda_{\mathrm{IV}}
=
\sigma_c^2.
\]
In this case, the IV estimator reduces to ordinary least squares.
\subsection{Recursive least squares and recursive IV implementation}
\label{recursive_ls}

As the number of data samples increases, constructing the full data
matrices may be impractical. Recursive updates address this by
incrementally updating the estimates and the corresponding inverse
matrices. Because the non-causal regressor
\[
\phi_{r,d}(k)
=
[u(k+d)^{\top}\;\cdots\;u(k-r)^{\top}]^{\top}
\]
contains the future inputs $u(k+1),\ldots,u(k+d)$, the update associated
with the output sample $y(k)$ can be performed only after time $k+d$.
Thus the implementation is online with a fixed delay of $d$ samples.
When $d=0$, it reduces to the usual causal online update.

For notational compactness, write
\[
\varphi_k\isdef \phi_{r,d}(k),
\qquad
z_k\isdef \phi_{c,r,d}(k).
\]
The vector $\varphi_k$ is the non-causal input regressor, while $z_k$ is
the corresponding instrument regressor. For ordinary least squares, the
standard RLS update rules \citep{azoury2001relative} apply directly to
$\varphi_k$. For each admissible output index
$k=r,\ldots,\ell-d$, define
\begin{equation}
g_{\mathrm{LS}}(k)
\isdef
\frac{
P_{k-1}^{\mathrm{LS}}\varphi_k
}{
\lambda_{\mathrm f}
+
\varphi_k^{\top}P_{k-1}^{\mathrm{LS}}\varphi_k
},
\label{eq:RLS-gain-corrected}
\end{equation}
where $P_k^{\mathrm{LS}}\in\mathbb{R}^{p\mu\times p\mu}$ and
$\lambda_{\mathrm f}\in(0,1]$ is the forgetting factor. The LS estimate
is updated as
\begin{align}
\hat{\theta}_{r,d,k}^{\mathrm{LS}}
&=
\hat{\theta}_{r,d,k-1}^{\mathrm{LS}}
+
\left[
y(k)
-
\hat{\theta}_{r,d,k-1}^{\mathrm{LS}}\varphi_k
\right]
g_{\mathrm{LS}}(k)^{\top},
\label{eq:RLS-thetaUpdate-corrected}
\\
P_k^{\mathrm{LS}}
&=
\frac{1}{\lambda_{\mathrm f}}
\left[
P_{k-1}^{\mathrm{LS}}
-
g_{\mathrm{LS}}(k)\varphi_k^{\top}P_{k-1}^{\mathrm{LS}}
\right].
\label{eq:RLS-P-corrected}
\end{align}
This recursion is memory efficient because it avoids explicitly storing
the full regressor matrix. However, for closed-loop data, LS may still
be biased because $\varphi_k$ can be correlated with the effective
regression error.

For closed-loop data, the recursive IV update is different from simply
substituting $z_k$ for $\varphi_k$ in the LS gain. The batch IV estimate
satisfies the normal equation
\[
\hat{\theta}_{r,d,\ell}^{\mathrm{IV}}
\left(
\Phi_{r,d,\ell}\Phi_{c,r,d,\ell}^{\top}
\right)
=
\Psi_{y,\ell}\Phi_{c,r,d,\ell}^{\top}.
\]
Thus, the recursive implementation must update the empirical
cross-covariance between the input regressor and the instrument
regressor. Define
\[
S_k
\isdef
\lambda_{\mathrm f}S_{k-1}
+
\varphi_k z_k^{\top},
\qquad
M_k
\isdef
\lambda_{\mathrm f}M_{k-1}
+
y(k)z_k^{\top}.
\]
The recursive IV estimate is then
\[
\hat{\theta}_{r,d,k}^{\mathrm{IV}}
=
M_kS_k^{-1}.
\]

Equivalently, let
\[
P_k^{\mathrm{IV}}\isdef S_k^{-1}.
\]
Using the Sherman-Morrison formula, define
\begin{align}
g_{\mathrm{IV}}(k)
&\isdef
\frac{
P_{k-1}^{\mathrm{IV}}\varphi_k
}{
\lambda_{\mathrm f}
+
z_k^{\top}P_{k-1}^{\mathrm{IV}}\varphi_k
},
\label{eq:RIV-column-gain}
\\
h_{\mathrm{IV}}(k)^{\top}
&\isdef
\frac{
z_k^{\top}P_{k-1}^{\mathrm{IV}}
}{
\lambda_{\mathrm f}
+
z_k^{\top}P_{k-1}^{\mathrm{IV}}\varphi_k
}.
\label{eq:RIV-row-gain}
\end{align}
Then the recursive IV update can be written as
\begin{align}
\hat{\theta}_{r,d,k}^{\mathrm{IV}}
&=
\hat{\theta}_{r,d,k-1}^{\mathrm{IV}}
+
\left[
y(k)
-
\hat{\theta}_{r,d,k-1}^{\mathrm{IV}}\varphi_k
\right]
h_{\mathrm{IV}}(k)^{\top},
\label{eq:RIV-thetaUpdate}
\\
P_k^{\mathrm{IV}}
&=
\frac{1}{\lambda_{\mathrm f}}
\left[
P_{k-1}^{\mathrm{IV}}
-
g_{\mathrm{IV}}(k)z_k^{\top}P_{k-1}^{\mathrm{IV}}
\right].
\label{eq:RIV-P-update}
\end{align}
Unlike the LS recursion, $P_k^{\mathrm{IV}}$ is the inverse of a
generally non-symmetric cross-covariance matrix. Therefore, the
recursion should be initialized with a regularized matrix, for example
\[
S_0=\eta I_{p\mu},
\qquad
P_0^{\mathrm{IV}}=\eta^{-1}I_{p\mu},
\qquad
\eta>0,
\]
and the denominator
\[
\lambda_{\mathrm f}
+
z_k^{\top}P_{k-1}^{\mathrm{IV}}\varphi_k
\]
must remain nonzero. This recursive IV implementation is the online
counterpart of the batch estimator in~\eqref{eq:IV_est}. The
finite-sample analysis below is stated for the batch IV estimator.

\section{Finite-Time Guarantees for Markov Parameter Estimation}
\label{sec:results}

We now derive a finite-sample error bound for the IV estimator in \eqref{eq:IV_est}. The goal is to quantify how accurately the finite block of Laurent coefficients $\theta_{r,d}$ can be estimated from one closed-loop trajectory. The unknown controller affects this error through the closed-loop state moments, the concentration of the feedback and instrument cross-covariance, and the instrument-strength parameter $\lambda_{\mathrm{IV}}$. The remaining terms in the bound capture the sample size, the FIR horizons, the process and measurement noise levels, and the stable and reverse-time unstable truncation
tails.

Under Assumption~\ref{ass:closedloop}, the closed-loop state has bounded second moments. We use the stable and unstable components of
the state-space decomposition and define
\begin{equation}
\begin{aligned}
\Gamma_{\mathrm{cl,s}}
&\isdef
\sup_{k\ge0}
\left\|
\mathbb{E}
[
x_{\mathrm{s}}(k)x_{\mathrm{s}}(k)^{\top}
]
\right\|
<\infty,
\\
\Gamma_{\mathrm{cl,u}}
&\isdef
\sup_{k\ge0}
\left\|
\mathbb{E}
[
x_{\mathrm{u}}(k)x_{\mathrm{u}}(k)^{\top}
]
\right\|
<\infty.
\end{aligned}
\label{eq:Gamma_cl}
\end{equation}
Here $x_{\mathrm{s}}(k)$ and $x_{\mathrm{u}}(k)$ are the stable and
unstable state components associated with the stable and unstable
realization of $G$.

\smallskip
\noindent
\textit{Why closed-loop stability enters.}
If an unstable plant is excited in open loop, the unstable state component can grow without bound. Then the non-causal truncation error $C_{\mathrm{u}}A_{\mathrm{u}}^{-d-1}x_{\mathrm{u}}(k+d+1)$ may have uncontrolled variance. Closed-loop stabilization is therefore
needed to keep the state moments bounded, while the non-causal
representation makes the multiplier $A_{\mathrm{u}}^{-d-1}$ contractive
as $d$ increases.

The stable and reverse-time unstable truncation scales are
\begin{equation}
\begin{aligned}
\sigma_{e,\mathrm{s}}^{\mathrm{cl}}
&\isdef
\Phi(A_{\mathrm{s}})
\|C_{\mathrm{s}}A_{\mathrm{s}}^{r}\|
\sqrt{
\frac{
r\Gamma_{\mathrm{cl,s}}
}{
1-\rho(A_{\mathrm{s}})^r
}
},
\\
\sigma_{e,\mathrm{u}}^{\mathrm{cl}}
&\isdef
\Phi(A_{\mathrm{u}}^{-1})
\|C_{\mathrm{u}}A_{\mathrm{u}}^{-d-1}\|
\sqrt{
\frac{
d\Gamma_{\mathrm{cl,u}}
}{
1-\rho(A_{\mathrm{u}}^{-1})^d
}
},
\end{aligned}
\label{eq:sigma_e_cl}
\end{equation}
where $\Phi(A) \isdef \sup_{\tau\ge0} \frac{\|A^\tau\|}{\rho(A)^{\tau/2}}$ measures transient amplification. The quantities in
\eqref{eq:sigma_e_cl} decay geometrically with $r$ and $d$, up to
transient-amplification and closed-loop covariance factors.

For a confidence level $\delta\in(0,1)$, define the logarithmic factor
used for the instrument covariance concentration:
\begin{equation}
\chi_N(\delta)
\isdef
\log^2\!\left(\frac{16\mu p}{\delta}\right)
\log^2\!\left(\frac{16Np}{\delta}\right).
\label{eq:chi_N_def}
\end{equation}
For the process-noise and instrument product, define
\begin{equation}
N_w(\delta)
\isdef
\kappa_w\mu (l+p)
L_{w,1}(\delta)^2
L_{w,2}(\delta)^2,
\label{eq:Nw_delta_def}
\end{equation}
where $L_{w,1}(\delta) \isdef \log\!\left(\frac{16\mu (l+p)}{\delta}\right),$ $L_{w,2}(\delta) \isdef \log\!\left(\frac{16N(l+p)}{\delta}\right)$, and $\kappa_w>0$ is an absolute constant. To keep the error terms compact, define
\begin{equation}
M_v(\delta)
\isdef
\mu p+m+\log\!\left(\frac{16}{\delta}\right),
\label{eq:Mv_def}
\end{equation}
and
\begin{equation}
\begin{aligned}
D_{\mathrm{s}}(N,r)
&\isdef
1+
\frac{mr}{N(1-\rho(A_{\mathrm{s}})^r)},
\\
M_{\mathrm{s}}(r,\delta)
&\isdef
rp+m+\log\!\left(\frac{16(r+1)}{\delta}\right),
\\
D_{\mathrm{u}}(N,d)
&\isdef
1+
\frac{md}{N(1-\rho(A_{\mathrm{u}}^{-1})^d)},
\\
M_{\mathrm{u}}(d,\delta)
&\isdef
dp+m+\log\!\left(\frac{16(d+1)}{\delta}\right).
\end{aligned}
\label{eq:tail_helper_terms}
\end{equation}
The four error scales used in the theorem are then
\begin{align}
\beta_w^{\mathrm{IV}}(\delta)
&\isdef
c_w\sigma_w\|\gamma_{r,d}\|
\max\left\{
\sqrt{N_w(\delta)},
\frac{N_w(\delta)}{\sqrt{N}}
\right\},
\label{eq:beta_w_corrected}
\\
\beta_v^{\mathrm{IV}}(\delta)
&\isdef
c_v\sigma_v
\sqrt{M_v(\delta)},
\label{eq:beta_v_corrected}
\\
\beta_{e,\mathrm{s}}^{\mathrm{cl}}(\delta)
&\isdef
c_{e,\mathrm{s}}
\sigma_{e,\mathrm{s}}^{\mathrm{cl}}
\sqrt{
D_{\mathrm{s}}(N,r)M_{\mathrm{s}}(r,\delta)
},
\label{eq:beta_es_corrected}
\\
\beta_{e,\mathrm{u}}^{\mathrm{cl}}(\delta)
&\isdef
c_{e,\mathrm{u}}
\sigma_{e,\mathrm{u}}^{\mathrm{cl}}
\sqrt{
D_{\mathrm{u}}(N,d)M_{\mathrm{u}}(d,\delta)
}.
\label{eq:beta_eu_corrected}
\end{align}
The constants
$c_w,c_v,c_{e,\mathrm{s}},c_{e,\mathrm{u}}$ are universal positive
constants.

\begin{assumption}
\rm
\label{ass:cl_concentration}
For the selected finite horizon $(r,d)$ and sample size $N$, the
closed-loop process induced by the unknown controller $\mathcal K$
satisfies the following empirical concentration event with probability
at least $1-\delta/4$:
\begin{align}
\left\|
\frac{1}{N}
\Phi_{f,r,d,\ell}
\Phi_{c,r,d,\ell}^{\top}
-
S_{fc}
\right\|
&\le
\frac{s_{\mathrm{IV}}}{4},
\label{eq:ass_fc_concentration}
\\
\left\|
\Psi_{e_{\mathrm{s}},\ell}
\Phi_{c,r,d,\ell}^{\top}
\right\|
&\le
\sigma_c\sqrt{N}\,
\beta_{e,\mathrm{s}}^{\mathrm{cl}}(\delta),
\label{eq:ass_es_concentration}
\\
\left\|
\Psi_{e_{\mathrm{u}},\ell}
\Phi_{c,r,d,\ell}^{\top}
\right\|
&\le
\sigma_c\sqrt{N}\,
\beta_{e,\mathrm{u}}^{\mathrm{cl}}(\delta).
\label{eq:ass_eu_concentration}
\end{align}
Here $S_{fc}$ is the population feedback and instrument
cross-covariance defined in~\eqref{eq:Ruc_decomp}. The matrices
$\Psi_{e_{\mathrm{s}},\ell}$ and $\Psi_{e_{\mathrm{u}},\ell}$ are
formed from the stable and reverse-time unstable truncation tails in
\eqref{truncationErrors2}.
\end{assumption}

\smallskip
\noindent
\textit{Interpretation of Assumption~\ref{ass:cl_concentration}.}
Assumption~\ref{ass:cl_concentration} is a closed-loop concentration
condition. It is used to control empirical cross-products involving the
feedback component and the truncation tails. The first inequality
requires the empirical feedback and instrument cross-covariance to
concentrate around its population value. The second and third
inequalities require the stable and reverse-time unstable truncation
tails to have controlled empirical correlation with the instrument.

For stable linear closed loops driven by Gaussian or sub-Gaussian
signals, such concentration bounds can be verified using standard
results for geometrically stable linear processes
\citep{simchowitz2018learning,sarkar2019near,sarkar2021finite,
tsiamis2019finite}, possibly with different universal constants.
Related finite-sample IV analyses for closed-loop state-space
identification appear in~\citep{szentpeteri2023non}. For general
nonlinear controllers, mean-square stability alone does not imply
empirical concentration. In that case,
Assumption~\ref{ass:cl_concentration} should be read as an explicit
regularity condition on the closed-loop process.

\begin{theorem}
\rm
\label{thm:IV_main}
Consider the system~\eqref{ssEqn} under the standing assumptions and
Assumption~\ref{ass:excitability}. Fix FIR horizons $r,d\ge0$, let
\[
\mu=r+d+1,
\qquad
\ell=N+r+d-1,
\]
and let $\delta\in(0,1)$. Suppose that the closed-loop empirical
concentration condition in Assumption~\ref{ass:cl_concentration}
holds for the selected horizon $(r,d)$ and sample size $N$. If
\begin{equation}
N
\ge
c_0\mu p\,\chi_N(\delta)
\max\left\{
1,
\frac{\sigma_c^2}{\lambda_{\mathrm{IV}}}
\right\},
\label{eq:sample_size_corrected}
\end{equation}
where $c_0>0$ is a universal constant, then, with probability at least
$1-\delta$,
\begin{equation}
\left\|
\hat{\theta}_{r,d,\ell}^{\mathrm{IV}}
-
\theta_{r,d}
\right\|
\le
\frac{
\beta_w^{\mathrm{IV}}(\delta)
+
\beta_{e,\mathrm{s}}^{\mathrm{cl}}(\delta)
+
\beta_{e,\mathrm{u}}^{\mathrm{cl}}(\delta)
+
\beta_v^{\mathrm{IV}}(\delta)
}{
\sqrt{\lambda_{\mathrm{IV}}N}
}.
\label{eq:IV_bound}
\end{equation}
\end{theorem}

\begin{prff}
See Appendix~\ref{theorem_proof}. \qed
\end{prff}

The bound in~\eqref{eq:IV_bound} separates four effects. The
process-noise term comes from the independent product of $w$ and $c$.
The measurement-noise term comes from the independent product of $v$
and $c$. The two truncation terms are controlled by the stable decay
$\rho(A_{\mathrm{s}})^r$ and the reverse-time unstable decay
$\rho(A_{\mathrm{u}}^{-1})^d$. The instrument quality enters through
\[
\lambda_{\mathrm{IV}}^{-1/2}
=
\frac{\sigma_c}{\sigma_{\min}(R_{uc})}.
\]
Thus small $\lambda_{\mathrm{IV}}$ corresponds to a weak instrument and
increases the estimation error.

\begin{corollary}
\rm
\label{cor:main}
Choose FIR orders
\begin{equation}
\begin{aligned}
r
&=
\mathcal{O}
\left(
\frac{1}{|\log\rho(A_{\mathrm{s}})|}
\log\left(\frac{N}{\varepsilon_0}\right)
\right),
\\
d
&=
\mathcal{O}
\left(
\frac{1}{|\log\rho(A_{\mathrm{u}}^{-1})|}
\log\left(\frac{N}{\varepsilon_0}\right)
\right),
\end{aligned}
\label{eq:rd_choice_corrected}
\end{equation}
where $\varepsilon_0\in(0,1)$ is a truncation tolerance. Then, with
probability at least $1-\delta$,
\begin{align}
\left\|
\hat{\theta}_{r,d,\ell}^{\mathrm{IV}}
-
\theta_{r,d}
\right\|
\le
C
\Bigg[
&
\frac{\sigma_v}{\sqrt{\lambda_{\mathrm{IV}}}}
\sqrt{
\frac{M_v(\delta)}{N}
}
\nonumber\\
&+
\frac{\sigma_w\|\gamma_{r,d}\|}
{\sqrt{\lambda_{\mathrm{IV}}}}
\max\left\{
\sqrt{\frac{N_w(\delta)}{N}},
\frac{N_w(\delta)}{N}
\right\}
\nonumber\\
&+
\frac{\varepsilon_0}{\sqrt{\lambda_{\mathrm{IV}}}}
\sqrt{
\frac{M_v(\delta)}{N}
}
\Bigg],
\label{eq:cor_bound}
\end{align}
where $C>0$ is a universal constant. If the error is measured with
respect to the infinite Laurent operator rather than the truncated FIR
parameter $\theta_{r,d}$, an additional deterministic approximation
term of order $\varepsilon_0$ must be added to the right-hand side.
\end{corollary}

\begin{prff}
See Appendix~\ref{proof:corollary}. \qed
\end{prff}

The order choice~\eqref{eq:rd_choice_corrected} is logarithmic in the
target accuracy $N/\varepsilon_0$ and inverse-logarithmic in the stable
and reverse-time unstable spectral radii. Near the unit circle,
$1/|\log\rho|$ behaves like the inverse stability gap, so the required
horizon grows when stable or reverse-time unstable modes are close to
the unit circle.

The finite-time result is best interpreted as follows: the FIR horizon needed to control truncation depends on the decay rates $\rho(A_{\mathrm{s}})$ and $\rho(A_{\mathrm{u}}^{-1})$. The constants in the Markov-parameter bound may also depend on transient amplification, closed-loop state moments, noise dimensions, and the instrument-conditioning parameter $\lambda_{\mathrm{IV}}$. Therefore, the result provides a finite-sample guarantee for estimating the Laurent/FIR Markov parameters from closed-loop data. Standard realization methods can be applied to the estimated Markov parameters when a state-space model is desired, but the present analysis focuses on the Markov-parameter estimation step.

\section{Experimental Results}
\vspace{-5pt}

In this section we consider numerical and real-world examples.

\begin{figure}[t]
\centering
\includegraphics[width=\columnwidth]{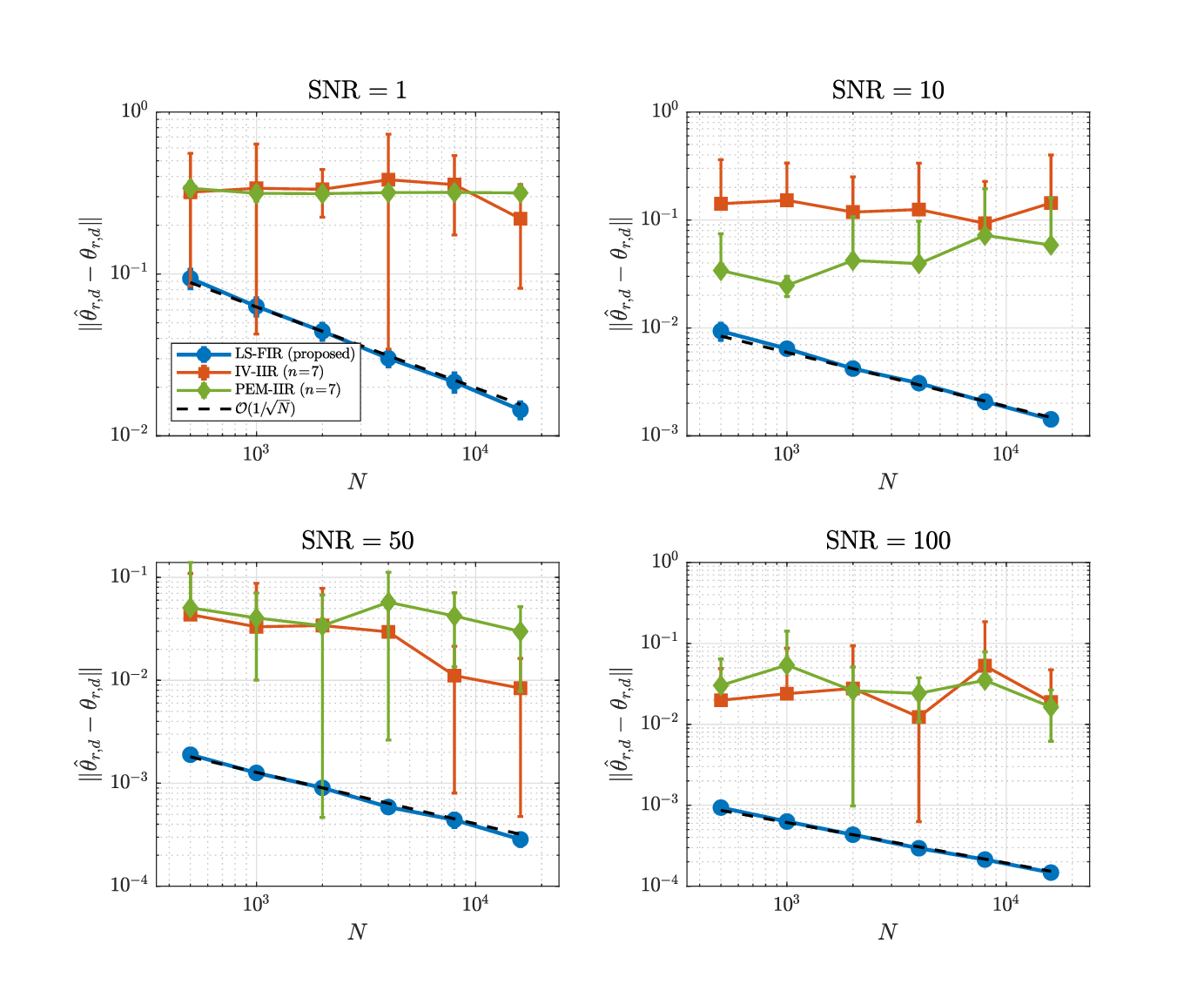}
\caption{Example~\ref{num_example}: Markov-parameter estimation error $\|\hat{\theta}_{r,d}-\theta_{r,d}\|$ versus $N$ for IV-FIR (proposed), IV-IIR ($n = 7$), and PEM-IIR ($n = 7$) at four SNR levels. The dashed line indicates the $\mathcal{O}\!\left(1/\sqrt{N}\right)$ rate.}
\label{fig:fig1}
\end{figure}

\begin{figure}[t]
\centering
\includegraphics[width=\columnwidth]{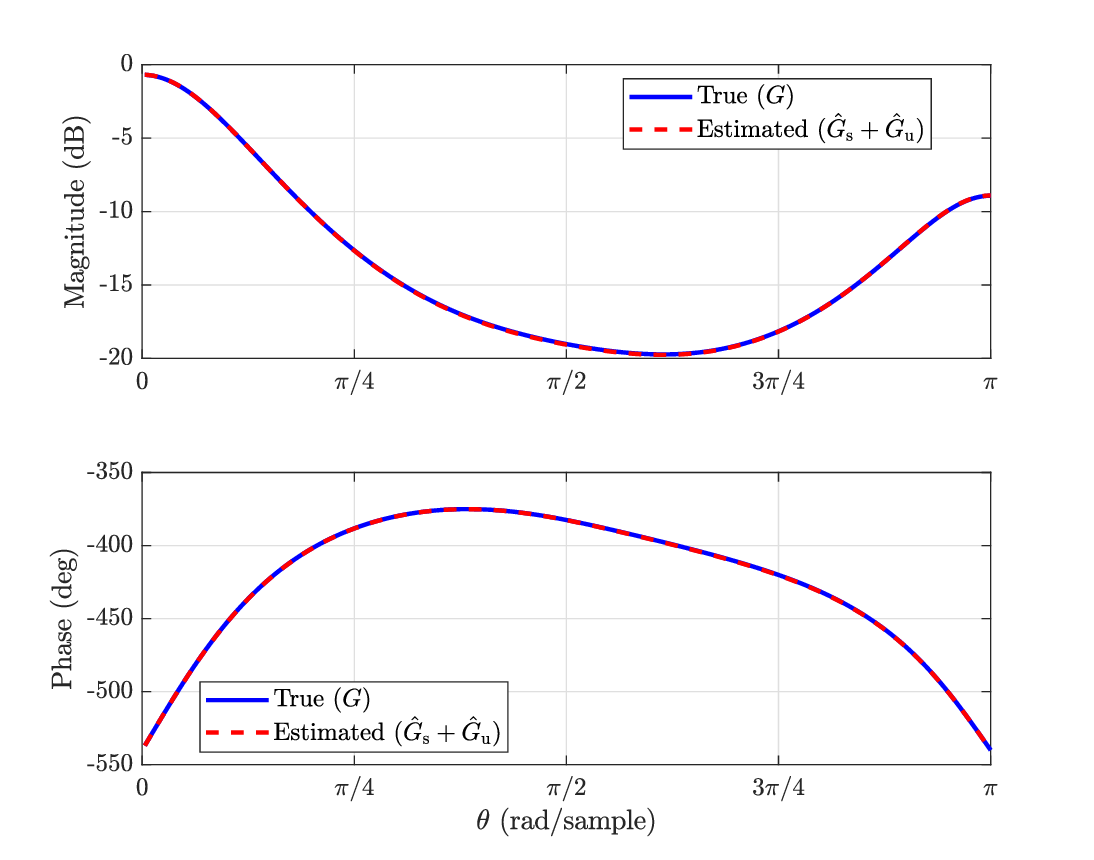}
\caption{Example~\ref{num_example}: Frequency response of the true
$G(z)$ and the reconstructed model
$\hat{G}_{\mathrm{s}}(z)+\hat{G}_{\mathrm{u}}(z)$ obtained via
Ho--Kalman from the estimated Markov parameters
(SNR $= 100$, $N = 16{,}000$).}
\label{fig:bode}
\end{figure}

\begin{examples}
\rm
\label{num_example}
Consider the seventh-order unstable transfer function
\begin{equation*}
G(z)=
\frac{(z^{2}\!+\!0.09)(z\!+\!0.2)(z\!-\!0.2)}{(z\!+\!0.6)(z\!+\!0.5)(z^{2}\!+\!0.16)(z\!-\!1.7)(z\!-\!1.6)(z\!-\!1.5)},
\end{equation*}
\textcolor{black}{with four stable poles ($-0.6$, $-0.5$, $\pm0.4j$) and three unstable poles ($1.5$, $1.6$, $1.7$), so $n_{\mathrm{s}}=4$, $n_{\mathrm{u}}=3$.} Consider also a corresponding state space realization $(A,B,C,D).$ We stabilize the plant with an LQR law ($Q=I_{7}$, $R=1$)and collect a single closed-loop trajectory: a Gaussian excitation $c(k)\sim\mathcal{N}(0,\sigma_{c}^{2})$ is added to the control input, so that $u(k)=-Kx(k)+c(k)$, and the measured output $y(k)=Cx(k)+Du(k)+v(k)$ is corrupted by zero-mean Gaussian measurement noise. The feedback gain $K$ is used only to generate the closed-loop data and is not used by the identification algorithm. The recorded signals are the measured input and output together with the injected excitation.

We benchmark three approaches:
\begin{enumerate}
\item[(i)] The proposed \textbf{non-causal IV-FIR} model with $r=d=25$ and $\mu=51$. The regressor $ \phi_{r,d}(k) = [u(k+d),\ldots,u(k),\ldots,u(k-r)]^{\top} $ contains future and past inputs, while the corresponding instrument regressor is $ \phi_{c,r,d}(k) = [c(k+d),\ldots,c(k),\ldots,c(k-r)]^{\top}.$ The Laurent/FIR coefficients are estimated using the batch IV estimator $ \hat{\theta}_{r,d,\ell}^{\mathrm{IV}} = \Psi_{y,\ell}\Phi_{c,r,d,\ell}^{\top} \left( \Phi_{r,d,\ell}\Phi_{c,r,d,\ell}^{\top} \right)^{-1}.$ No knowledge of the stabilizing controller or exact system order is required; only the FIR horizons \(r\) and \(d\) are chosen.
\item[(ii)] A \textbf{seventh-order IIR model via IV}
\citep{szentpeteri2023non}: MATLAB's \texttt{iv4} is used with ARX
structure $n_{a}\!=\!n_{b}\!=\!7$; the instruments are simulated
outputs from an auxiliary ARX model estimated in a preliminary step \citep{ljung1998system}.
\item[(iii)] A \textbf{seventh-order IIR model via PEM}, representing predictor-based approaches \citep{lale2020logarithmic,lale2021adaptive,jones2022closed,lale2021finite,lee2020non}. MATLAB's \texttt{pem} is used with state-space model order $n\!=\!7$.
\end{enumerate}
Note that both IIR baselines are given the true system order $n=7$, whereas the proposed IV-FIR approach only requires the chosen FIR horizons $r$ and $d$.

Figure~\ref{fig:fig1} plots $\|\theta_{r,d}-\hat{\theta}_{r,d,\ell}^{\mathrm{IV}}\|$ versus $N$ for SNR$\,\in\{1,10,50,100\}$. Across all noise levels, the proposed non-causal IV-FIR method attains the smallest Markov-parameter error and exhibits a smooth $\mathcal{O}(1/\sqrt{N})$ learning curve. The IIR baselines remain larger and exhibit higher variance, since they rely on finite-order IIR model fitting and, in the PEM case, a nonconvex optimization. In contrast, the proposed FIR estimator is a finite-dimensional linear IV estimator for the Laurent/FIR coefficients.

We then split $\hat{\theta}_{r,d,\ell}$ into causal and non-causal parts and apply the Ho-Kalman algorithm to recover separate stable and unstable realizations, that is, $(\hat{A}_{\mathrm{s}},\hat{B}_{\mathrm{s}},\hat{C}_{\mathrm{s}})$ and $(\hat{A}_{\mathrm{u}},\hat{B}_{\mathrm{u}},\hat{C}_{\mathrm{u}})$, respectively, which can be combined to reconstruct the transfer function $\hat{G}(z)=\hat{G}_{\mathrm{s}}(z)+\hat{G}_{\mathrm{u}}(z)$. Figure~\ref{fig:bode} compares the frequency response of the estimated transfer function $\hat{G}$ with the true transfer function $G$. The close agreement in both magnitude and phase supports the quality of the estimated Laurent/FIR Markov parameters.
\end{examples}

\begin{examples}
\rm
\label{ex:hair_dryer}
We next consider input-output data from a hair dryer system DaISy~\citep{de1997daisy}, where the input is the heater voltage and
the output is the air temperature. This benchmark is useful because no
stability information is provided to the estimator a priori. The purpose
of this example is to illustrate that the non-causal FIR parameterization
can be used without knowing in advance whether the underlying dynamics
are stable or unstable.

We estimate a non-causal FIR model with $r=d=25$ using the recursive least-squares implementation described in Section~\ref{recursive_ls}. Since the regressor contains future inputs, the recursive update is available with a fixed $d$-sample delay. We compare the resulting RLS-FIR estimate with IV-IIR and PEM-IIR baselines. Figure~\ref{fig:dryer_ex_samples} shows that the non-causal RLS-FIR estimator converges faster and reaches a lower steady-state prediction error than the IIR baselines.

Figure~\ref{fig:dryer_ex} shows the measured and predicted outputs, together with the estimated Laurent/FIR coefficients. The estimated negative-lag coefficients are close to zero, while the positive-lag coefficients capture the dominant response. This behavior is consistent with a stable thermal system: if the dynamics are stable, the non-causal part of the Laurent representation should be negligible. Thus, although the estimator is not given stability information, the estimated coefficients reveal that the data are well described by a stable causal
response.

\begin{figure}[t]
\centering
\includegraphics[width=0.8\columnwidth]{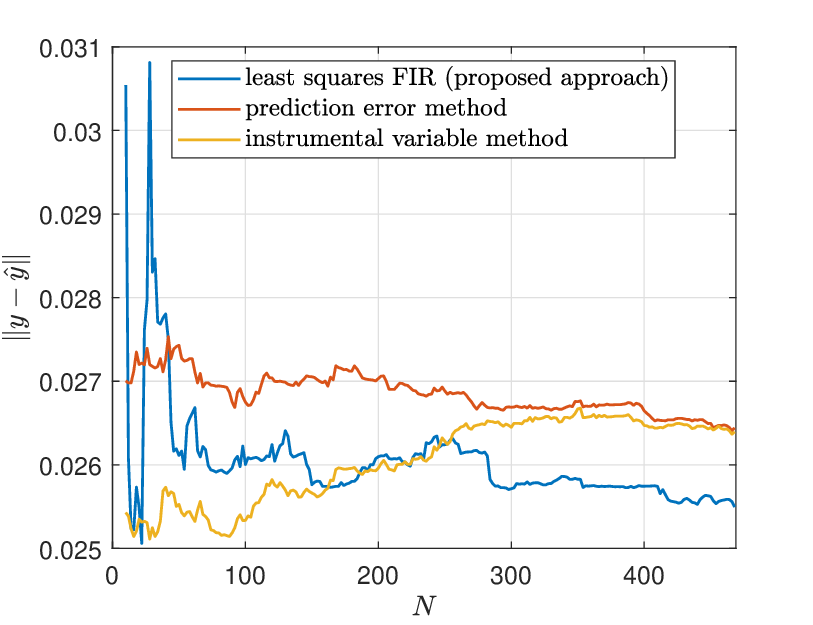}
\caption{Example~\ref{ex:hair_dryer}: Parameter estimation errors for RLS-FIR, IV-IIR, and PEM-IIR.}
\label{fig:dryer_ex_samples}
\end{figure}

\begin{figure}[t]
\centering
\includegraphics[width=\columnwidth]{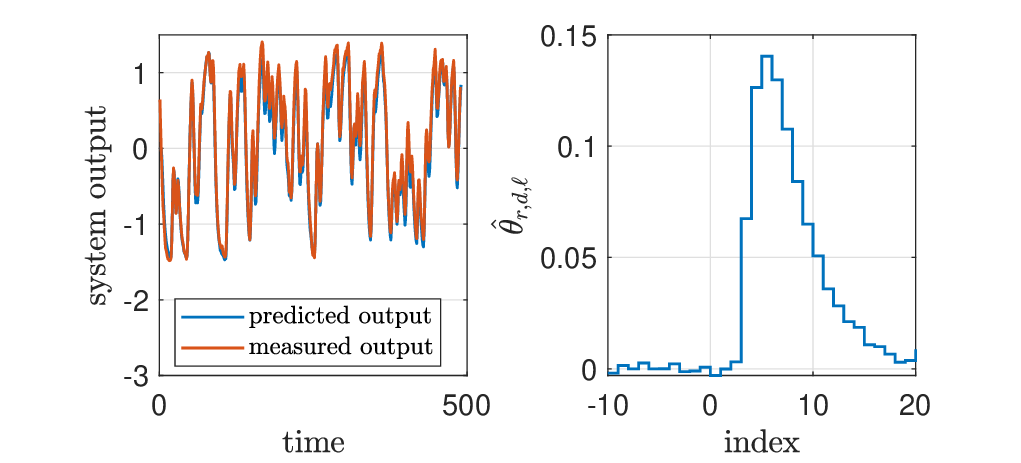}
\caption{Example~\ref{ex:hair_dryer}: Predicted vs.\ measured output and estimated Laurent/FIR coefficients.}
\label{fig:dryer_ex}
\end{figure}

\end{examples}

\begin{figure}[t]
\centering
\includegraphics[width=\columnwidth]{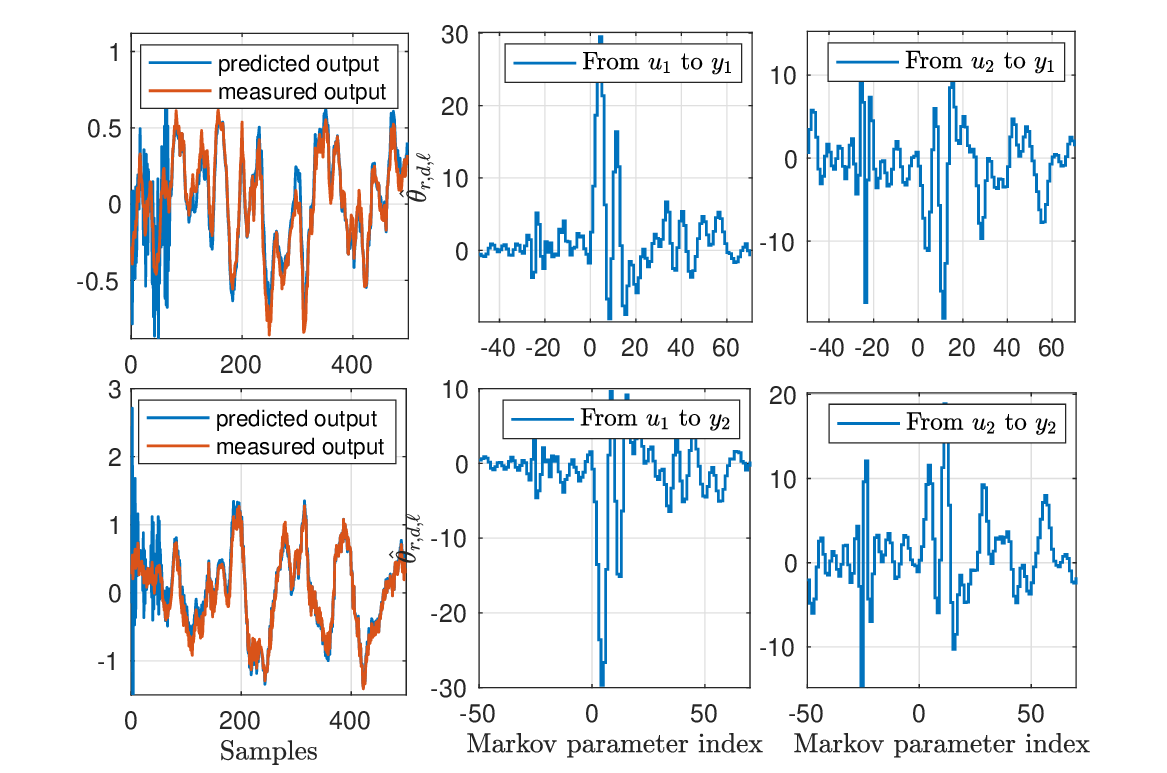} \caption{Example~\ref{ex:cd_arm_main}: Predicted and measured outputs with estimated Laurent/FIR coefficients for the CD-player arm benchmark.}
\label{fig:fig2}
\end{figure}

\begin{examples}
\label{ex:cd_arm_main}
\rm
The DaISy~\citep{de1997daisy} CD-player arm benchmark is a $2\times2$ MIMO system with actuator force inputs and laser-based position outputs. The data were collected in closed loop. This example is used as a real-data illustration of the non-causal FIR parameterization in a MIMO setting, rather than as a direct validation of the finite-sample IV theorem.

We estimate a non-causal FIR model with $r=d=50$ using the recursive least-squares implementation of Section~\ref{recursive_ls}. Since the regressor contains future inputs, the recursive update is available with a fixed $d$-sample delay. Figure~\ref{fig:fig2} shows the measured and predicted outputs, together with the estimated Laurent/FIR coefficients. The predicted outputs are obtained directly from the estimated non-causal FIR model.

The left panel of Figure~\ref{fig:fig2} shows strong agreement between the predicted and measured outputs. The estimated coefficients in the right panel contain significant values at both positive and negative lags. The positive-lag coefficients represent the causal stable component, while the negative-lag coefficients represent the non-causal component associated with reverse-time stable dynamics. Thus, although some standard preprocessing may suggest a stable model, the estimated Laurent/FIR coefficients indicate that a non-causal component is useful for explaining the closed-loop MIMO data. This supports the role of the proposed representation as a diagnostic and modeling tool for systems whose stability structure is not known a priori.
\end{examples}

\begin{figure}[t]
\centering
\includegraphics[width=\columnwidth]{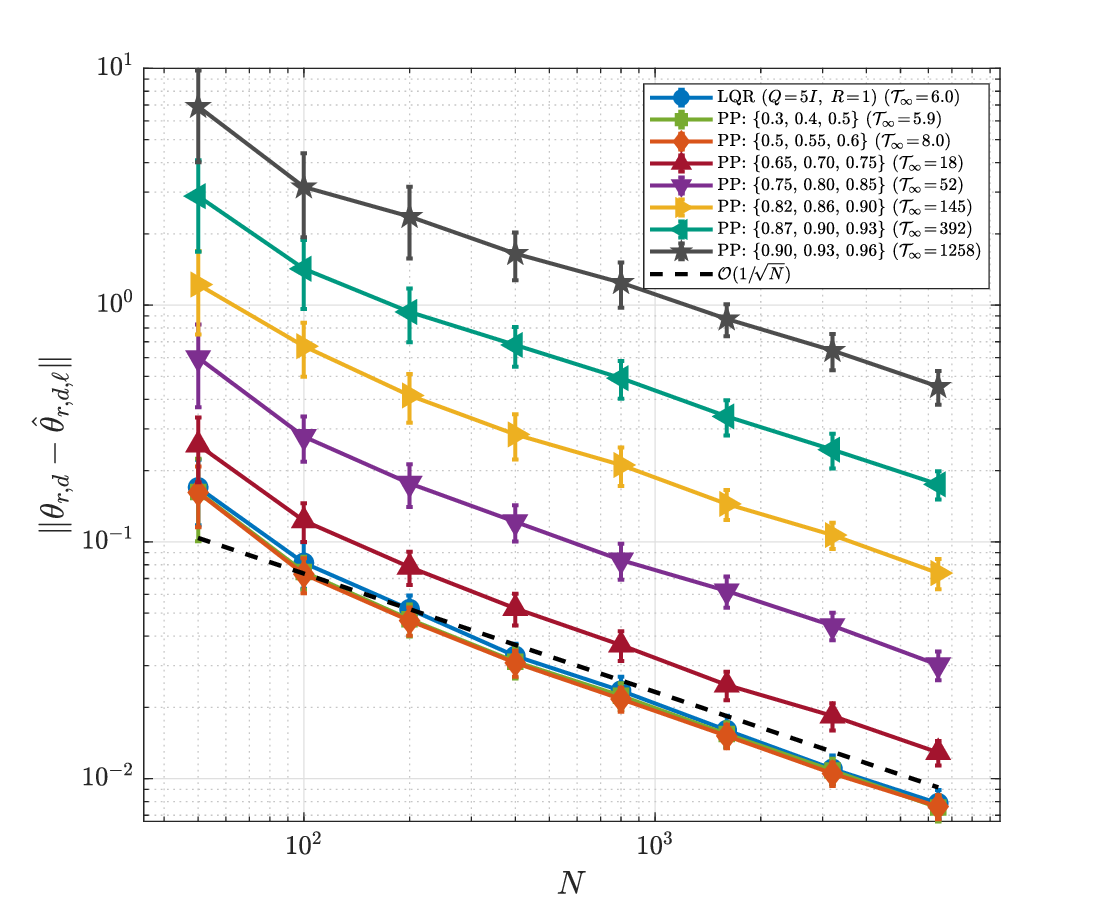}
\caption{Example~\ref{ex:controller_effect}: Markov-parameter error of the non-causal IV-FIR estimator for different stabilizing controllers. Larger $\mathcal{T}_{\infty}$ corresponds to slower decay of the feedback response from the injected excitation to the feedback signal, which leads to weaker instrument conditioning and larger sample complexity.}
\label{fig:controller_effect}
\end{figure}

\begin{examples}
\rm
\label{ex:controller_effect}
In this example, we examine how the choice of stabilizing controller affects the sample complexity of closed-loop identification. As discussed in Section~\ref{sec:IV}, the controller influences the IV estimator through the feedback and instrument cross-covariance. In the linear case, the quantity $\mathcal{T}_{\infty}$ provides a conservative measure of how strongly past injected excitations are recirculated through the feedback signal.

Consider the third-order plant $A=\mathrm{diag}(0.3,\,1.5,\,2.0)$,  $B=[1\;\;1\;\;1]^\top$, $C=[1\;\;1\;\;1]$ with two unstable poles. We design eight stabilizing linear controllers: one LQR law and seven pole-placement designs. The closed-loop spectral radii range from $\rho_{\mathrm{cl}}=0.50$ to $\rho_{\mathrm{cl}}=0.96$, producing $\mathcal{T}_{\infty}$ values
ranging from approximately $6$ to $1258$. For each controller, we run
$80$ simulation trials at SNR$\,=20$ using $r=d=20$ and
$N\in\{50,\ldots,6400\}$. The Markov parameters are estimated using the
non-causal IV-FIR estimator with the injected excitation $c$ as the
instrument.

Figure~\ref{fig:controller_effect} shows the 
Markov-parameter error 
$\|\hat\theta_{r,d}-\theta_{r,d}\|$ versus the number of samples $N$. Two observations stand out. First, all tested controllers exhibit the predicted $\mathcal{O}(1/\sqrt{N})$ decay, consistent with the finite-sample analysis. Second, the vertical gap between the curves is strongly correlated with $\mathcal{T}_{\infty}$. As the closed-loop poles move closer to the unit circle, the impulse response from the injected excitation $c$ to the feedback signal $f$ decays more slowly. This increases $\mathcal{T}_{\infty}$ and is associated with weaker instrument conditioning, reflected by a smaller effective $\lambda_{\mathrm{IV}}$.

The barely stabilizing controller ($\mathcal{T}_{\infty}\approx1258$) requires substantially more samples than the LQR design ($\mathcal{T}_{\infty}\approx6$) to reach comparable accuracy. Since all controllers in this experiment are linear and strictly causal, the finite-horizon population cross-covariance $R_{uc}$ retains the triangular structure described in Section~\ref{sec:IV}. Thus large $\mathcal{T}_{\infty}$ does not make the IV construction invalid; rather, it worsens the conditioning and therefore increases the sample complexity. This experiment highlights that controller design is an important degree of freedom in closed-loop identification: a controller may stabilize the plant while still leading to weak instrument conditioning and slower learning.
\end{examples}

\begin{examples}
\rm
\label{example_deep_baselines}

Following the benchmark of \citep{pillonetto2025deep}, consider the
SISO plant
\begin{equation}
G(z)
=
\frac{z-0.1}{(z-0.4)(z+0.2)(z+0.5)} .
\end{equation}
This system is open-loop stable. Therefore, a purely causal FIR model
would be sufficient if the stability information were known in advance.
Here, however, we intentionally apply the full non-causal FIR
parameterization with $r=d=10$ to illustrate that the proposed
representation does not require prior knowledge of the stability
structure. In this stable case, the estimated negative-lag coefficients
$\hat H_{-1},\ldots,\hat H_{-d}$ are close to zero, so the non-causal
model effectively reduces to a causal FIR model.

Since this benchmark is open-loop stable and no feedback-induced
correlation is present, the IV estimator reduces to ordinary least
squares. We therefore report the LS implementation of the FIR estimator
for this example. This experiment is intended to illustrate the
computational simplicity and predictive accuracy of the FIR
parameterization, rather than the closed-loop IV bias-removal effect.

We compare the proposed FIR estimator with six deep generative
state-space models that learn latent dynamics through variational
inference and recurrent architectures: STORN
\citep{liu2023improved,bayer2014learning}, VAE-RNN
\citep{gedon2021deep,chung2015recurrent}, VRNN-Gauss
\citep{chung2015recurrent}, VRNN-Gauss-I, VRNN-GMM
\citep{chung2015recurrent}, and VRNN-GMM-I. Figure~\ref{fig:baseline_rnn}
shows the test-set RMSE versus the number of training samples $N$.
The FIR estimator achieves the lowest RMSE with substantially fewer
parameters and negligible training time compared with the deep
state-space baselines, as reported in Table~\ref{tab:baseline_rnn}.
This highlights the practical advantage of the proposed representation
when a compact linear model is appropriate.

\begin{figure}[t]
\centering
\includegraphics[width=\columnwidth]{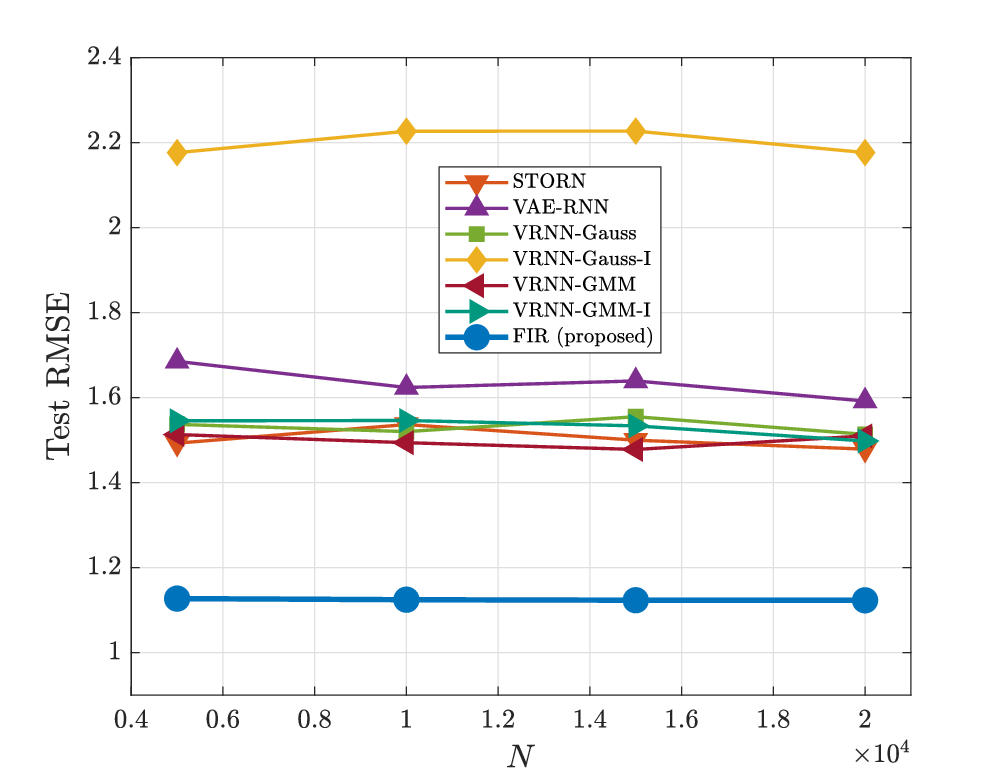}
\caption{Example~\ref{example_deep_baselines}: Test-set RMSE versus
training samples for the FIR estimator and deep state-space models.}
\label{fig:baseline_rnn}
\end{figure}

\begin{table*}[t]
\centering
\caption{Training time and model complexity.}
\label{tab:baseline_rnn}
\begin{tabular}{@{}lrr@{}}
    \toprule
    \textbf{Model} & \textbf{Time (s)} & \textbf{\# Params} \\
    \midrule
    STORN \citep{liu2023improved,bayer2014learning} & 3107.5 & 120492 \\
    VAE-RNN \citep{gedon2021deep}                   & 3227.0 & 81292  \\
    VRNN-Gauss \citep{chung2015recurrent}           & 3206.6 & 105862 \\
    VRNN-Gauss-I                                    & 1971.5 & 90242  \\
    VRNN-GMM \citep{chung2015recurrent}             & 3143.3 & 101815 \\
    VRNN-GMM-I                                      & 3236.1 & 101815 \\
    \midrule
    FIR estimator                                   & \textbf{0.003} & \textbf{21} \\
    \bottomrule
\end{tabular}
\end{table*}

\end{examples}

\section{Conclusion}
This paper introduced a non-causal FIR framework for finite-time identification of stable and unstable LTI systems from a single closed-loop trajectory. The Laurent/FIR representation captures unstable dynamics through reverse-time stable coefficients, which keeps the input and process-noise terms controlled by stable decay rates rather than by growing unstable dynamics. To address closed-loop bias, we used the injected excitation as an instrumental variable, without requiring knowledge of the stabilizing controller. Under explicit instrument-strength and closed-loop concentration conditions, we established an $\mathcal{O}(N^{-1/2})$ Markov-parameter error bound, up to logarithmic factors and truncation terms. The analysis also shows how the controller affects sample complexity through instrument conditioning, and the numerical results support the predicted finite-sample behavior.

\bibliography{newBibFile.bib}           

\newpage
\appendix

\noindent
This appendix contains two parts. Section~\ref{app:decoupling} gives
the stable and unstable decomposition used in the Laurent/FIR
representation and truncation analysis. Section~\ref{theorem_proof}
proves Theorem~\ref{thm:IV_main} and Corollary~\ref{cor:main}.

\section{Stable and Unstable Decomposition}
\label{app:decoupling}

We decompose the original realization into stable and unstable
subsystems. This decomposition justifies the two-sided Laurent
representation and controls the stable and reverse-time unstable
truncation errors.

Since $A$ has no eigenvalues on the unit circle, there exists a real
Schur decomposition
\[
A_t
\isdef
V^{\top}AV
=
\begin{bmatrix}
A_{11} & A_{12}\\
0 & A_{22}
\end{bmatrix},
\]
where $V$ is orthogonal, all eigenvalues of $A_{11}$ lie inside the
unit disk, and all eigenvalues of $A_{22}$ lie outside the unit disk.
Let $n_{\mathrm{s}}$ and $n_{\mathrm{u}}$ denote the corresponding
dimensions.

The off-diagonal coupling $A_{12}$ can be removed by a second
similarity transformation. Let
\[
W
=
\begin{bmatrix}
I_{n_{\mathrm{s}}} & S\\
0 & I_{n_{\mathrm{u}}}
\end{bmatrix},
\]
where $S$ solves the Sylvester equation
\[
A_{11}S-SA_{22}+A_{12}=0.
\]
This equation has a unique solution because the spectra of $A_{11}$ and
$A_{22}$ are disjoint. Define
\[
x_d(k)
\isdef
W^{-1}V^{\top}x(k)
=
\begin{bmatrix}
x_{\mathrm{s}}(k)\\
x_{\mathrm{u}}(k)
\end{bmatrix}.
\]
Then the dynamics become

\begin{equation}
\begin{aligned}
\begin{bmatrix}
x_{\mathrm{s}}(k+1)\\
x_{\mathrm{u}}(k+1)
\end{bmatrix}
&=
\begin{bmatrix}
A_{\mathrm{s}} & 0\\
0 & A_{\mathrm{u}}
\end{bmatrix}
\begin{bmatrix}
x_{\mathrm{s}}(k)\\
x_{\mathrm{u}}(k)
\end{bmatrix}
+
\begin{bmatrix}
B_{\mathrm{s}}\\
B_{\mathrm{u}}
\end{bmatrix}
u(k) \\
& +
\begin{bmatrix}
B_{\mathrm{s},w}\\
B_{\mathrm{u},w}
\end{bmatrix}
w(k),
\\
y(k)
&=
\begin{bmatrix}
C_{\mathrm{s}} & C_{\mathrm{u}}
\end{bmatrix}
\begin{bmatrix}
x_{\mathrm{s}}(k)\\
x_{\mathrm{u}}(k)
\end{bmatrix}
+
Du(k)+v(k),
\end{aligned}
\label{eq:decoupled_appendix}
\end{equation}
where
\[
\rho(A_{\mathrm{s}})<1,
\qquad
\rho(A_{\mathrm{u}}^{-1})<1.
\]

Under Assumption~\ref{ass:closedloop}, the closed-loop state has
uniformly bounded second moments. Since the above transformations are
finite-dimensional and invertible, the stable and unstable components
also have bounded second moments:
\[
\Gamma_{\mathrm{cl,s}}
=
\sup_{k\ge0}
\left\|
\mathbb E[
x_{\mathrm{s}}(k)x_{\mathrm{s}}(k)^{\top}]
\right\|
<\infty,
\]
and
\[
\Gamma_{\mathrm{cl,u}}
=
\sup_{k\ge0}
\left\|
\mathbb E[
x_{\mathrm{u}}(k)x_{\mathrm{u}}(k)^{\top}]
\right\|
<\infty .
\]

The stable truncation tail in~\eqref{truncationErrors2} is
\[
e_{\mathrm{s}}(k)
=
C_{\mathrm{s}}A_{\mathrm{s}}^{r}x_{\mathrm{s}}(k-r).
\]
Therefore,
\[
\left\|
\mathbb E[
e_{\mathrm{s}}(k)e_{\mathrm{s}}(k)^{\top}]
\right\|
\le
\|C_{\mathrm{s}}A_{\mathrm{s}}^{r}\|^{2}
\Gamma_{\mathrm{cl,s}}.
\]
Since $\rho(A_{\mathrm{s}})<1$, this term decays geometrically with
$r$, up to transient-amplification constants.

For the unstable component, the reverse-time representation is governed
by $A_{\mathrm{u}}^{-1}$. Indeed,
\[
x_{\mathrm{u}}(k)
=
A_{\mathrm{u}}^{-1}x_{\mathrm{u}}(k+1)
-
A_{\mathrm{u}}^{-1}B_{\mathrm{u}}u(k)
-
A_{\mathrm{u}}^{-1}B_{\mathrm{u},w}w(k).
\]
This reverse-time transition is stable because
$\rho(A_{\mathrm{u}}^{-1})<1$. The non-causal truncation tail is
\[
e_{\mathrm{u}}(k)
=
C_{\mathrm{u}}A_{\mathrm{u}}^{-d-1}x_{\mathrm{u}}(k+d+1).
\]
Hence,
\[
\left\|
\mathbb E[
e_{\mathrm{u}}(k)e_{\mathrm{u}}(k)^{\top}]
\right\|
\le
\|C_{\mathrm{u}}A_{\mathrm{u}}^{-d-1}\|^{2}
\Gamma_{\mathrm{cl,u}}.
\]
Since $\rho(A_{\mathrm{u}}^{-1})<1$, this term decays geometrically
with the preview horizon $d$.

\section{Proof of Theorem~\ref{thm:IV_main} and Corollary~\ref{cor:main}}
\label{theorem_proof}

We prove the finite-sample bound for the instrumental-variable
estimator. The proof keeps separate the population instrument strength
\[
s_{\mathrm{IV}}
\isdef
\sigma_{\min}(R_{uc})
\]
from the normalized conditioning parameter
\[
\lambda_{\mathrm{IV}}
\isdef
\frac{s_{\mathrm{IV}}^2}{\sigma_c^2}.
\]
This normalization is useful because the empirical inverse
cross-covariance contributes a factor $1/s_{\mathrm{IV}}$, while
products involving the instrument carry the scale $\sigma_c$. Thus
\[
\frac{\sigma_c}{s_{\mathrm{IV}}}
=
\frac{1}{\sqrt{\lambda_{\mathrm{IV}}}}.
\]

For readability, write
\[
\Phi
\isdef
\Phi_{r,d,\ell},
\qquad
\Phi_c
\isdef
\Phi_{c,r,d,\ell},
\qquad
\Phi_f
\isdef
\Phi_{f,r,d,\ell}.
\]
Recall that $\ell=N+r+d-1$ and $\mu=r+d+1$.

Define the aggregate disturbance matrix
\begin{equation}
E_{r,d,\ell}
\isdef
\gamma_{r,d}\Phi_{w,r,d,\ell}
+
\Psi_{e_{r,d},\ell}
+
\Psi_{v,\ell}.
\label{eq:E_matrix_appendix}
\end{equation}
From the truncated non-causal FIR representation,
\begin{equation}
\Psi_{y,\ell}
=
\theta_{r,d}\Phi
+
E_{r,d,\ell}.
\label{eq:data_identity_appendix}
\end{equation}
Substituting~\eqref{eq:data_identity_appendix} into the IV estimator
gives
\begin{align}
\hat{\theta}_{r,d,\ell}^{\mathrm{IV}}
&=
\left(
\theta_{r,d}\Phi
+
E_{r,d,\ell}
\right)
\Phi_c^{\top}
\left(
\Phi\Phi_c^{\top}
\right)^{-1}
\nonumber\\
&=
\theta_{r,d}
+
E_{r,d,\ell}
\Phi_c^{\top}
\left(
\Phi\Phi_c^{\top}
\right)^{-1}.
\end{align}
Therefore,
\begin{equation}
\hat{\theta}_{r,d,\ell}^{\mathrm{IV}}
-
\theta_{r,d}
=
E_{r,d,\ell}
\Phi_c^{\top}
\left(
\Phi\Phi_c^{\top}
\right)^{-1}.
\label{eq:IV_error_identity_appendix}
\end{equation}
Taking spectral norms yields
\begin{align}
\left\|
\hat{\theta}_{r,d,\ell}^{\mathrm{IV}}
-
\theta_{r,d}
\right\|
\le
\Big(
&
\|\gamma_{r,d}\|
\,
\|\Phi_{w,r,d,\ell}\Phi_c^{\top}\|
\nonumber\\
&+
\|\Psi_{e_{r,d},\ell}\Phi_c^{\top}\|
+
\|\Psi_{v,\ell}\Phi_c^{\top}\|
\Big)
\nonumber\\
&\times
\left\|
\left(
\Phi\Phi_c^{\top}
\right)^{-1}
\right\|.
\label{eq:four_terms_appendix}
\end{align}

\subsection{Cross-covariance concentration}

We first control the inverse factor in
\eqref{eq:four_terms_appendix}. Since
\[
\Phi=\Phi_f+\Phi_c,
\]
we have
\begin{align}
\frac{1}{N}\Phi\Phi_c^{\top}
-
R_{uc}
=
&
\left(
\frac{1}{N}\Phi_c\Phi_c^{\top}
-
\sigma_c^2I_{p\mu}
\right)
\nonumber\\
&+
\left(
\frac{1}{N}\Phi_f\Phi_c^{\top}
-
S_{fc}
\right),
\label{eq:cross_cov_decomposition_appendix}
\end{align}
where $S_{fc}$ is defined in~\eqref{eq:Ruc_decomp}.

By standard finite-sample covariance concentration for Gaussian block
regressors, as used in non-asymptotic system identification analyses
\citep{oymak2019non,simchowitz2018learning,sarkar2021finite}, there
exists a universal constant $c_0>0$ such that, if
\begin{equation}
N
\ge
c_0\mu p\,\chi_N(\delta)
\max
\left\{
1,
\frac{\sigma_c^2}{\lambda_{\mathrm{IV}}}
\right\},
\label{eq:sample_size_cross_cov_appendix}
\end{equation}
then, with probability at least $1-\delta/4$,
\begin{equation}
\left\|
\frac{1}{N}\Phi_c\Phi_c^{\top}
-
\sigma_c^2I_{p\mu}
\right\|
\le
\frac{s_{\mathrm{IV}}}{4}.
\label{eq:instrument_cov_concentration_appendix}
\end{equation}

The second term in~\eqref{eq:cross_cov_decomposition_appendix}
contains the feedback component. By
Assumption~\ref{ass:cl_concentration}, with probability at least
$1-\delta/4$,
\begin{equation}
\left\|
\frac{1}{N}\Phi_f\Phi_c^{\top}
-
S_{fc}
\right\|
\le
\frac{s_{\mathrm{IV}}}{4}.
\label{eq:feedback_cov_concentration_appendix}
\end{equation}
Combining~\eqref{eq:cross_cov_decomposition_appendix},
\eqref{eq:instrument_cov_concentration_appendix}, and
\eqref{eq:feedback_cov_concentration_appendix} gives
\begin{equation}
\left\|
\frac{1}{N}\Phi\Phi_c^{\top}
-
R_{uc}
\right\|
\le
\frac{s_{\mathrm{IV}}}{2}.
\label{eq:total_cov_deviation_appendix}
\end{equation}
By Weyl's inequality,
\[
\sigma_{\min}
\left(
\frac{1}{N}\Phi\Phi_c^{\top}
\right)
\ge
s_{\mathrm{IV}}
-
\frac{s_{\mathrm{IV}}}{2}
=
\frac{s_{\mathrm{IV}}}{2}.
\]
Equivalently,
\begin{equation}
\sigma_{\min}
\left(
\Phi\Phi_c^{\top}
\right)
\ge
\frac{Ns_{\mathrm{IV}}}{2},
\qquad
\left\|
\left(
\Phi\Phi_c^{\top}
\right)^{-1}
\right\|
\le
\frac{2}{Ns_{\mathrm{IV}}}.
\label{eq:inverse_cross_cov_appendix}
\end{equation}

\smallskip
\noindent
\textit{Connection with prior concentration tools.}
The covariance bound above is the point where the IV setting differs
from the open-loop least-squares setting. In open loop, the relevant
Gram matrix involves an independent input regressor with itself. Here,
the empirical matrix involves the feedback-affected input regressor
against the injected excitation. The decomposition
$\Phi=\Phi_f+\Phi_c$ separates the directly excited part from the
feedback part, and the latter is controlled through
Assumption~\ref{ass:cl_concentration}.

\subsection{Process-noise term}

The process-noise regressor $\Phi_{w,r,d,\ell}$ and the instrument
matrix $\Phi_c$ are formed from mutually independent Gaussian sequences.
Therefore, standard Gaussian product concentration
\citep{oymak2019non,tsiamis2019finite} gives, with probability at least
$1-\delta/4$,
\begin{equation}
\left\|
\Phi_{w,r,d,\ell}
\Phi_c^{\top}
\right\|
\le
c_w\sigma_w\sigma_c
\max
\left\{
\sqrt{N_w(\delta)N},
N_w(\delta)
\right\}.
\label{eq:process_product_bound_appendix}
\end{equation}
Multiplying by $\|\gamma_{r,d}\|$ and using the definition of
$\beta_w^{\mathrm{IV}}(\delta)$ gives
\begin{equation}
\|\gamma_{r,d}\|
\left\|
\Phi_{w,r,d,\ell}
\Phi_c^{\top}
\right\|
\le
\sigma_c\sqrt{N}\,
\beta_w^{\mathrm{IV}}(\delta).
\label{eq:process_term_appendix}
\end{equation}

\subsection{Measurement-noise term}

Similarly, the measurement-noise matrix $\Psi_{v,\ell}$ and the
instrument matrix $\Phi_c$ are formed from mutually independent
Gaussian sequences. Applying the same Gaussian product concentration
argument gives, with probability at least $1-\delta/4$,
\begin{equation}
\left\|
\Psi_{v,\ell}
\Phi_c^{\top}
\right\|
\le
c_v\sigma_v\sigma_c
\sqrt{
N M_v(\delta)
}.
\label{eq:measurement_product_bound_appendix}
\end{equation}
By the definition of $\beta_v^{\mathrm{IV}}(\delta)$, this can be
written as
\begin{equation}
\left\|
\Psi_{v,\ell}
\Phi_c^{\top}
\right\|
\le
\sigma_c\sqrt{N}\,
\beta_v^{\mathrm{IV}}(\delta).
\label{eq:measurement_term_appendix}
\end{equation}

\subsection{Truncation-error term}

The truncation-error matrix decomposes as
\[
\Psi_{e_{r,d},\ell}
=
\Psi_{e_{\mathrm{s}},\ell}
+
\Psi_{e_{\mathrm{u}},\ell},
\]
where $\Psi_{e_{\mathrm{s}},\ell}$ is formed from the stable tail
\[
C_{\mathrm{s}}A_{\mathrm{s}}^{r}x_{\mathrm{s}}(k-r),
\]
and $\Psi_{e_{\mathrm{u}},\ell}$ is formed from the reverse-time
unstable tail
\[
C_{\mathrm{u}}A_{\mathrm{u}}^{-d-1}x_{\mathrm{u}}(k+d+1).
\]
By Assumption~\ref{ass:cl_concentration},
\begin{equation}
\left\|
\Psi_{e_{\mathrm{s}},\ell}
\Phi_c^{\top}
\right\|
\le
\sigma_c\sqrt{N}\,
\beta_{e,\mathrm{s}}^{\mathrm{cl}}(\delta),
\label{eq:stable_tail_product_appendix}
\end{equation}
and
\begin{equation}
\left\|
\Psi_{e_{\mathrm{u}},\ell}
\Phi_c^{\top}
\right\|
\le
\sigma_c\sqrt{N}\,
\beta_{e,\mathrm{u}}^{\mathrm{cl}}(\delta).
\label{eq:unstable_tail_product_appendix}
\end{equation}
Therefore,
\begin{equation}
\left\|
\Psi_{e_{r,d},\ell}
\Phi_c^{\top}
\right\|
\le
\sigma_c\sqrt{N}
\left(
\beta_{e,\mathrm{s}}^{\mathrm{cl}}(\delta)
+
\beta_{e,\mathrm{u}}^{\mathrm{cl}}(\delta)
\right).
\label{eq:tail_total_product_appendix}
\end{equation}

\subsection{Assembling the theorem}

Define
\begin{equation}
\begin{aligned}
\mathcal B_{r,d}(\delta)
\isdef&
\beta_w^{\mathrm{IV}}(\delta)
+
\beta_{e,\mathrm{s}}^{\mathrm{cl}}(\delta)
+
\beta_{e,\mathrm{u}}^{\mathrm{cl}}(\delta)
+
\beta_v^{\mathrm{IV}}(\delta).
\end{aligned}
\label{eq:B_total_appendix}
\end{equation}
On the intersection of the concentration events
\eqref{eq:inverse_cross_cov_appendix},
\eqref{eq:process_term_appendix},
\eqref{eq:measurement_term_appendix}, and
\eqref{eq:tail_total_product_appendix}, substituting into
\eqref{eq:four_terms_appendix} gives
\begin{align}
\left\|
\hat{\theta}_{r,d,\ell}^{\mathrm{IV}}
-
\theta_{r,d}
\right\|
&\le
\sigma_c\sqrt{N}\,
\mathcal B_{r,d}(\delta)
\frac{2}{Ns_{\mathrm{IV}}}
\nonumber\\
&=
\frac{2\mathcal B_{r,d}(\delta)}
{\sqrt{\lambda_{\mathrm{IV}}N}} .
\label{eq:assembled_bound_appendix}
\end{align}
The universal factor $2$ is absorbed into the constants defining the
$\beta$ terms. A union bound over the four events gives probability at
least $1-\delta$. Therefore,
\begin{equation}
\left\|
\hat{\theta}_{r,d,\ell}^{\mathrm{IV}}
-
\theta_{r,d}
\right\|
\le
\frac{
\mathcal B_{r,d}(\delta)
}{
\sqrt{\lambda_{\mathrm{IV}}N}
},
\end{equation}
which proves Theorem~\ref{thm:IV_main}. \qed

\subsection{Proof of Corollary~\ref{cor:main}}
\label{proof:corollary}

We now choose the FIR horizons $r$ and $d$ so that the deterministic
stable and reverse-time unstable Laurent tails are small.

For the stable part,
\begin{equation}
\sum_{i=r+1}^{\infty}
\left\|
C_{\mathrm{s}}A_{\mathrm{s}}^{i-1}B_{\mathrm{s}}
\right\|
\le
\frac{
\Phi(A_{\mathrm{s}})
\|C_{\mathrm{s}}\|
\|B_{\mathrm{s}}\|
\rho(A_{\mathrm{s}})^r
}{
1-\rho(A_{\mathrm{s}})
}.
\label{eq:stable_tail_sum_appendix}
\end{equation}
For the reverse-time unstable part,
\begin{equation}
\sum_{i=d+1}^{\infty}
\left\|
C_{\mathrm{u}}A_{\mathrm{u}}^{-i-1}B_{\mathrm{u}}
\right\|
\le
\frac{
\Phi(A_{\mathrm{u}}^{-1})
\|C_{\mathrm{u}}A_{\mathrm{u}}^{-2}\|
\|B_{\mathrm{u}}\|
\rho(A_{\mathrm{u}}^{-1})^d
}{
1-\rho(A_{\mathrm{u}}^{-1})
}.
\label{eq:unstable_tail_sum_appendix}
\end{equation}
Therefore, choosing
\begin{equation}
\begin{aligned}
r
&=
\mathcal{O}
\left(
\frac{1}{|\log \rho(A_{\mathrm{s}})|}
\log\frac{N}{\varepsilon_0}
\right),
\\
d
&=
\mathcal{O}
\left(
\frac{1}{|\log \rho(A_{\mathrm{u}}^{-1})|}
\log\frac{N}{\varepsilon_0}
\right)
\end{aligned}
\label{eq:rd_choice_appendix}
\end{equation}
makes the stable and reverse-time unstable truncation contributions of
order $\varepsilon_0$, up to transient-amplification and decomposition
constants.

With this choice, the truncation terms in
Theorem~\ref{thm:IV_main} satisfy
\begin{equation}
\beta_{e,\mathrm{s}}^{\mathrm{cl}}(\delta)
+
\beta_{e,\mathrm{u}}^{\mathrm{cl}}(\delta)
\le
C_e\varepsilon_0
\sqrt{
M_v(\delta)
},
\label{eq:tail_beta_cor_appendix}
\end{equation}
up to the logarithmic factors already contained in the theorem. Here
$C_e>0$ depends on transient amplification and closed-loop covariance
bounds.

Substituting~\eqref{eq:tail_beta_cor_appendix} and the definitions of
$\beta_w^{\mathrm{IV}}(\delta)$ and
$\beta_v^{\mathrm{IV}}(\delta)$ into
Theorem~\ref{thm:IV_main} gives
\begin{align}
\left\|
\hat{\theta}_{r,d,\ell}^{\mathrm{IV}}
-
\theta_{r,d}
\right\|
\le
C
\Bigg[
&
\frac{\sigma_v}{\sqrt{\lambda_{\mathrm{IV}}}}
\sqrt{
\frac{M_v(\delta)}{N}
}
\nonumber\\
&+
\frac{\sigma_w\|\gamma_{r,d}\|}
{\sqrt{\lambda_{\mathrm{IV}}}}
\max
\left\{
\sqrt{\frac{N_w(\delta)}{N}},
\frac{N_w(\delta)}{N}
\right\}
\nonumber\\
&+
\frac{\varepsilon_0}{\sqrt{\lambda_{\mathrm{IV}}}}
\sqrt{
\frac{M_v(\delta)}{N}
}
\Bigg],
\label{eq:corollary_bound_appendix}
\end{align}
for a universal constant $C>0$.

If the comparison is made against the infinite Laurent operator rather
than the truncated parameter $\theta_{r,d}$, then the deterministic
truncation approximation contributes an additional additive term of
order $\varepsilon_0$. This proves Corollary~\ref{cor:main}. \qed
\noindent

\end{document}